\documentclass[11pt,letterpaper,onecolumn]{IEEEtran}
\pdfoutput =1

\usepackage{amsmath,epsfig,amssymb,verbatim,amsopn,color}
\usepackage{cite,xspace}
\usepackage{array,algorithm,algorithmic}
\usepackage{array}
\usepackage{multirow,array}%
\usepackage{multicol}%
\usepackage{caption}%
\usepackage{enumerate}
\usepackage{subfig} 
\usepackage{url}

\ifCLASSOPTIONpeerreview
\renewcommand{\baselinestretch}{1.95}
\fi

\ifCLASSOPTIONonecolumn
    \renewcommand{\baselinestretch}{1.75}
    \usepackage[margin=25mm]{geometry}
\fi

\makeatletter
\renewcommand*{\@opargbegintheorem}[3]{\trivlist
  \item[\hskip \labelsep{\itshape #1\ #2}] {\itshape (#3):} {\normalfont}}
\makeatother

\newcommand{\dv}[1]{\boldsymbol{#1}}

\newcommand{\rr}{r_{\textrm{o}}}

\newcommand{\pis}{P_{\textrm{iso}}(\rr)}
\newcommand{\pcon}{P_{\textrm{1-con}}(\rr)}
\newcommand{\pkcon}{P_{\textrm{k-con}}(\rr)}

\newcommand{\figref}[1]{Fig.\,\ref{#1}}

\newcommand{\secref}[1]{Section\,\ref{#1}}
\newcommand{\rba}[1]{B_{#1}(\dv{u};\rr)}
\newcommand{\rca}[1]{C_{#1}(\dv{u};\rr)}

\usepackage{relsize}

\newenvironment{xitemize}{%
    \itemize
}{%
    \enditemize
}

\newtheorem{definition}{Definition}

\newtheorem{remark}{Remark}

\newcommand{\AuthorOne}{Zubair~Khalid, {\em{Member, IEEE}}}
\newcommand{\AuthorTwo}{Salman~Durrani, {\em{Senior Member, IEEE}}}
\newcommand{\AuthorThree}{Jing Guo, {\em{Student Member, IEEE}}}

\newcommand{\ThankOne}{The authors are with the Research School of Engineering, College of Engineering and Computer Science, The
Australian National University, Canberra, ACT 0200, Australia.
Emails: \{zubair.khalid, salman.durrani, jing.guo\}@anu.edu.au}

\begin{document}

\title{A Tractable Framework for Exact Probability of Node Isolation and Minimum Node Degree Distribution in Finite Multi-hop Networks}
\author{\authorblockN{\AuthorOne, \AuthorTwo \: and \AuthorThree \thanks{\ThankOne}}}

\maketitle

%
\begin{abstract}
This paper presents a tractable analytical framework
for the exact calculation of the probability of node isolation and
the minimum node degree distribution when $N$ sensor nodes are
independently and uniformly distributed inside a finite square
region. The proposed framework can accurately account for the
boundary effects by partitioning the square into subregions, based
on the transmission range and the node location. We show that for
each subregion, the probability that a random node falls inside a
disk centered at an arbitrary node located in that subregion can be
expressed analytically in closed-form. Using the results for the
different subregions, we obtain the exact probability of node
isolation and minimum node degree distribution that serves as an
upper bound for the probability of $k$-connectivity. Our theoretical
framework is validated by comparison with the simulation results and
shows that the minimum node degree distribution serves as a tight
upper bound for probability of $k$-connectivity. The proposed
framework provides a very useful tool to accurately account for the
boundary effects in the design of finite wireless networks.
\end{abstract}


\ifCLASSOPTIONpeerreview
    \newpage
\fi

\section{Introduction}
Wireless multi-hop networks, also refereed to as wireless sensor
networks and wireless ad hoc networks, consist of a group of sensor
nodes deployed over a finite
region~\cite{Gupta-2001,hong2002load,Puccinelli:2005,Cui-06,Li:2007}.
The nodes operate in a decentralized manner without the need of any
fixed infrastructure, i.e., the nodes communicate with each other
via a single-hop wireless path~(if they are in range) or via a
multi-hop wireless path. In most of the applications, such wireless
networks are formed by distributing a finite (small) number of nodes
in a finite area, which is typically assumed to be a square
region~\cite{Fan2007,Fabbri:2008,Andrews-2010,Peng:2010}.

Connectivity is a basic requirement for the planning and effective
operation of wireless multi-hop
networks~\cite{Mao-2012c,Baccelli-2009}. The $k$-connectivity is a
most general notion  of connectivity and an important
characteristic of wireless multi-hop
networks~\cite{bettstetter-2002,Peng:2010,Khalid:2013:Ausctw}. The
network being $k$-connected ensures that there exists at least $k$
\emph{independent} multi-hop paths between any two nodes. In other
words, $k$-connected network would still be $1$-connected if $(k-1)$
nodes forming the network fail. The \textit{probability of node
isolation}, defined as the probability that a randomly selected node
has no connections to any other nodes, plays a key role in
determining the overall network
connectivity~(1-connectivity)~\cite{bettstetter-2002,Bettstetter:2004}.
The \textit{minimum node degree distribution}, which is the
probability that each node in the network has at least $k$
neighbours, is crucial in determining the $k$-connectivity of the
network~\cite{bettstetter-2002}.

For large-scale wireless sensor networks, assuming Poisson
distributed nodes in an infinite area, the connectivity properties
such as probability of isolation, average node degree,
$k$-connectivity have been well
studied~\cite{Penrose-1997,Gupta-1998,bettstetter-2002,Ling:2007,Miorandi-2008,Zhou:2007:Ausctw,Zhou-2008,Ng:2013}.
When the node locations follow an infinite homogeneous Poisson point
process and assuming all nodes have the same transmission range, it
has been shown that the network becomes $k$-connected with high
probability~(close to $1$) at the same time the minimum node degree
of the network approaches $k$~\cite{Penrose-1997}. This fact was
used to approximate the probability of $k$-connectivity by the minimum
node degree distribution in \cite{bettstetter-2002}. It was also
used to determine the asymptotic value of the minimum transmission
range for $k$-connectivity for a uniform distribution of nodes in a
unit square and disk~\cite{Peng:2010}.

\subsection{Related Work}

Since many practical multi-hop networks are formed by distributing a
finite number of nodes in a finite area, there has been an
increasing interest to model and determine the connectivity
properties in finite multi-hop
networks~\cite{bettstetter-2002,Bettstetter:2004,Jia-2006,Fabbri-2008,Peng:2010,Srinivasa-2010,Eslami-2012,Mao-2012,Coon-2012}.
This is also due to the fact, established earlier in
\cite{bettstetter-2002,Bettstetter:2004} and recently in
\cite{Eslami-2012}, that the asymptotic connectivity results for
large-scale networks provide an extremely poor approximation for
finite wireless networks. This poor approximation is due to the
boundary effects experienced by the nodes near the borders of the
finite region over which the nodes are deployed. Since the nodes
located close to the physical boundaries of the network have a
limited coverage area, they have a greater probability of isolation.
Therefore, the boundary effects play an important role in
determining the overall network connectivity.

Different  approaches have been used in the literature, to try to
model the boundary effects including (i) using geometrical
probability~\cite{Mathai-1999} and dividing the square  region into
smaller subregions to facilitate asymptotic analysis of the
transmission range for $k$-connectivity~\cite{Peng:2010,Jia-2006}
and to find mean node degree in different
subregions~\cite{Nze-2011}, (ii) using a cluster expansion approach
and decomposing the boundary effects into corners and edges to yield
high density approximations~\cite{Coon-2012} and (iii) using a
deterministic grid deployment of nodes in a finite
area~\cite{Rajagopalan-2009} to approximate the boundary effects
with random deployment of nodes~\cite{Eslami-2012}. The above
approaches provide bounds, rather than exact results, for the
probability of node isolation and/or probability of connectivity.
For a wireless network deployed over a finite area, the existing results
for $k$-connectivity and minimum node degree are
asymptotic~(infinite $N$)~\cite{Peng:2005,Peng:2010}. An attempt was
made in \cite{Ling:2007} to study the minimum node degree and
$k$-connectivity by circumventing modeling of the boundary effects
but the results were shown to be valid for large density~(number of
nodes) only. Therefore, it is still largely an open research problem
to characterize the boundary effects and to find general frameworks
for deriving the exact results for the probability of node isolation and
the minimum node degree distribution, when a finite number of nodes are
independently and uniformly distributed inside a finite region.

\subsection{Contributions}

In the above context, we address the following open questions in
this paper for a wireless network of $N$ nodes, which are uniformly
distributed over a square region:

\begin{enumerate}[Q1]
\item How can we accurately account for the boundary effects to determine the exact probability of node isolation?
\item How can we incorporate the boundary effects to find the minimum node degree distribution?
\end{enumerate}

In this paper, addressing the above two open questions, we present a
tractable analytical framework for the exact calculation of the
probability of node isolation and the minimum node degree distribution
in finite wireless multi-hop networks, when $N$ nodes are
independently and uniformly distributed in a square region. Our
proposed framework partitions the square into unequal subregions,
based on the transmission range and the location of an arbitrary
node. Using geometrical probability, we show that for each
subregion, the probability that a random node falls inside a disk
centered at an arbitrary node located in that subregion can be
expressed analytically in closed-form. This framework accurately
models the boundary effects and leads to an exact expression for the
probability of node isolation and the minimum node degree distribution,
which can be easily evaluated numerically. We show that the minimum
node degree distribution can be used as an upper bound for the
probability of $k$-connectivity.

Since the $k$-connectivity depends on the number of nodes deployed
over the finite region and the transmission range of each
node~\cite{Peng:2005,Peng:2010}, the transmission range must be
large enough to ensure that the network is connected but small enough to
minimize the power consumption at each node and interference between
nodes~\cite{Santi:2001,bettstetter-2002}, which in turn maximizes
the network capacity. This fundamental trade-off between the network
connectivity and the network capacity leads to the following network
design question:
\begin{enumerate}[Q3]
\item  Given a network of $N$ nodes distributed over a square region, what
is the minimum transmission range such that a network is connected
with a high probability or alternatively, what is the minimum number
of nodes for a given transmission range such that the network is
connected?
\end{enumerate}
Addressing this network design problem, we show through an example
how the proposed framework can be used to determine the minimum
transmission range required for the network to be connected with
high probability.

The rest of the paper is organized as follows. The system model,
problem formulation and connectivity properties of a wireless
network are presented in \secref{sec:system_model}. The proposed
framework to evaluate the probability of node isolation and the minimum node
degree distribution is provided in \secref{sec:framework}. The
boundary effects in the different regions formed with the change in
transmission range are presented in Section~IV. The validation of
the proposed framework via simulation results and the design example
are presented in \secref{sec:results}. Finally,
\secref{sec:conclusions} concludes the paper.

\section{System Model and Problem Formulation}\label{sec:system_model}

\subsection{Distribution of Nodes and Node Transmission Model}
Consider $N$ nodes which are uniformly and independently distributed inside a square region $\mathcal{R} \in \mathbb{R}^2$, where $\mathbb{R}^2$ denotes the two dimensional Euclidean domain. Let $S_\ell$ and $V_\ell$, for $\ell\in\{1,2,3,4\}$, denote the side and vertex of the square, respectively, which are numbered in an anticlockwise direction. Without loss of generality, we assume that the first vertex $V_1$ of the square is located at the origin $(0,0)$ and we consider a unit square region defined as
\begin{align}\label{Eq:sqaure_region_def}
\mathcal{R} = \{\dv{u} = (x,y) \in \mathbb{R}^2|\, 0 \leq x \leq
1,\, 0 \leq y \leq 1\}.
\end{align}

Let $\dv{u} = (x,y)$ denote the position of an arbitrary node inside the square $\mathcal{R}$. The node distribution probability density function (PDF) can be expressed as
\begin{align}\label{Eq:pdf_uniform}
f_{\mathbf{U}}(\dv{u}) =   \begin{cases}
            1 \,\, & \dv{u}\in \mathcal{R},  \\
             0\,\,& \dv{u}\in  \mathbb{R}^2 \backslash \mathcal{R}.   \\
                    \end{cases}
\end{align}

We define $|\mathcal{R}| = \int_{\mathcal{R}}ds(\dv{u})$
as a measure of the physical area of the square region, where
$ds(\dv{u}) = dxdy$ and the integration is performed over the two
dimensional square region $\mathcal{R}$. Note that $|\mathcal{R}|=1$
since we assume a unit square.

We assume that each sensor node has a fixed transmission range $\rr$
and the \emph{coverage region} of a node located at $\dv{u}$ is then
a disk $\mathcal{O}(\dv{u};\rr)$ of radius $\rr$ centered at the
node. Note that the \textit{coverage area}
$|\mathcal{O}(\dv{u};\rr)| =\pi\rr^2$. The number of nodes inside
the coverage area of a certain node are termed as its neighbors.

\subsection{Connectivity Properties}

In this subsection, we define the key connectivity properties of a
multi-hop network, which are considered in this paper.

\begin{definition}[Conditional Probability of Connectivity]
Let the cumulative distribution function~(CDF) $F(\dv{u};\rr)$
denote the conditional probability of connectivity that a randomly
placed node according to uniform probability density function (PDF) is connected to a node located
at $\dv{u}$. Mathematically,
\begin{align}\label{Eq:CDF_general}
F(\dv{u};\rr)\,&\triangleq \,
{|\mathcal{O}(\dv{u};\rr)\cap\mathcal{R}|}.
\end{align}
\end{definition}

\begin{definition}[Probability of Node Isolation]
Let $\pis$ denote the probability of node isolation that any node
in the network is isolated. Assuming that the probability of node
isolation is independent for each node, the probability that a given
node at $\dv{u}$ is isolated is given by $\left(1-
F(\dv{u};\rr)\right)^{N-1}$, which can be averaged over all possible
locations to evaluate $\pis$ as
\begin{align}\label{Eq:piso}
\pis\,&\triangleq\, \int_{\mathbb{R}^2}  \left(1-
F(\dv{u};\rr)\right)^{N-1} f_{\mathbf{U}}(\dv{u}) ds(\dv{u})
\nonumber \\ &= \int_\mathcal{R} \left(1- F(\dv{u};\rr)\right)^{N-1}
ds(\dv{u}).
\end{align}
\end{definition}

\begin{definition}[Minimum Node Degree]
For a uniform distribution of $N$ nodes in a square region, define
the minimum node degree as the minimum of number of neighbors of
\emph{any} node in the region. Let the discrete random variable $D$
denote the minimum node degree. The associated PDF, termed as the
minimum node degree distribution is given by
\begin{align}\label{Eq:min_node_degree}
f_D(k; \rr) &= \textrm{P}(D=k) \triangleq  \bigg(1-
\sum_{d=0}^{k-1}\binom{N-1}{d} \times \nonumber \\ &\int_\mathcal{R} \left( F(\dv{u};\rr)
\right)^{d} \left(1- F(\dv{u};\rr) \right)^{N-d-1} ds(\dv{u})
\bigg)^N.
\end{align}
The details of the formulation of $f_D(k;\rr)$ are provided in
Appendix~A.
\end{definition}

\begin{definition}[$1$-connected network]
A network of $N$ nodes is said to be $1$-connected~(or connected) if
there exists at least one path between any pair of randomly chosen
nodes.
\end{definition}

\begin{definition}[$k$-connected network]
A network of $N$ nodes is said to be
$k$-connected~($k=1,2,\hdots,\,N-1$) if there exist at least $k$
mutually independent paths between any pair of randomly chosen nodes.
In other words, a network is $k$-connected if the network stays
$1$-connected with the removal of any $(k-1)$ nodes. Let $\pkcon$
denote the probability that the network of $N$ nodes~(each with
transmission range $\rr$) is $k$-connected.
\end{definition}

Next we examine the relation between probability $\pkcon$ and the
minimum node degree distribution $f_D(k;\rr)$.
Penrose~\cite{Penrose-1997} presented in his work on graph theory
that a random network for large enough number of nodes, becomes
$k$-connected at the same instant it acheives the minimum node
degree $k$ with high probability, that is, $f_D(k;\rr)$ serves as an
upper bound on $\pkcon$, which gets tighter as both $f_D(k;\rr)$ and
$\pkcon$ approach one or the number of nodes approaches infinity.
Mathematically, we can express this as
\begin{align}
f_D(k;\rr) & \geq \pkcon, \nonumber \\
f_D(k;\rr) & = \pkcon, \quad \pkcon \rightarrow 1. \label{Eq:pkcon_bound}
\end{align}
\noindent We note that the minimum node degree distribution is of
fundamental importance~\cite{bettstetter-2002} as (i) it determines the
connectivity of the network~($\pcon$), (ii) takes into account the failure of
the nodes and (iii) also determines the minimum node degree of the
network~($\pkcon$). Using \eqref{Eq:piso} and
\eqref{Eq:min_node_degree}, we also note the relationship between
$\pis$ and $f_D(k;\rr)$: $f_D(1;\rr) = (1-\pis)^{N}$. Since
$f_D(1;\rr)$ denotes the probability that each node has at least one
neighbor, it has been also referred to as the probability of no isolated
node in the literature~\cite{bettstetter-2002,Bettstetter:2004}.

\subsection{Problem Statement}

There are two key challenges in evaluating the probability of node
isolation $\pis$ in \eqref{Eq:piso} and the minimum node degree
distribution $f_D(k;\rr)$ in \eqref{Eq:min_node_degree}. The
\textit{first challenge} is to find the CDF in
\eqref{Eq:CDF_general}, which requires the evaluation of the overlap
area $|\mathcal{O}(\dv{u};\rr)\cap\mathcal{R}|$.
In~\cite{Fabbri-2008}, it is proposed to find this intersection area
using polar coordinates and dividing the square into different
radial regions. However, due to the dependance between the polar
radius and the polar angle, this approach does not lead to
closed-form solutions. In~\cite{Zubair-2012}, an alternative
approach is presented for finding the intersection area by first
finding the area of circular segments formed outside the sides and
vertices and then subtracting from the area of the disk. This
approach leads to closed-form solutions and is adopted in this work.

The \textit{second challenge} is to average the CDF given in
\eqref{Eq:CDF_general} over the square in order to determine the
probability of node isolation $\pis$ in \eqref{Eq:piso} and the
minimum node degree distribution $f_D(k;\rr)$ in
\eqref{Eq:min_node_degree}. $F(\dv{u};\rr)$ is a function of both
the node location $\dv{u}$ and the transmission range $\rr$. For a
unit square, if $\rr \geq \sqrt{2}$, then the disk
$\mathcal{O}(\dv{u};\rr)$ will cover the whole square $\mathcal{R}$
and hence $F(\dv{u};\rr)=1$, irrespective of the node location. For
intermediate values of the node range $0 \leq \rr \leq \sqrt{2}$,
both $\dv{u}$ and $\rr$ need to be taken into account in determining
$F(\dv{u};\rr)$. This adds further complexity to the task of
evaluating \eqref{Eq:piso} and \eqref{Eq:min_node_degree}. A
tractable exact solution to this problem is presented in the next
section.

\ifCLASSOPTIONonecolumn
\begin{figure*}[t]
 \centering
    \includegraphics[width=0.65\textwidth]{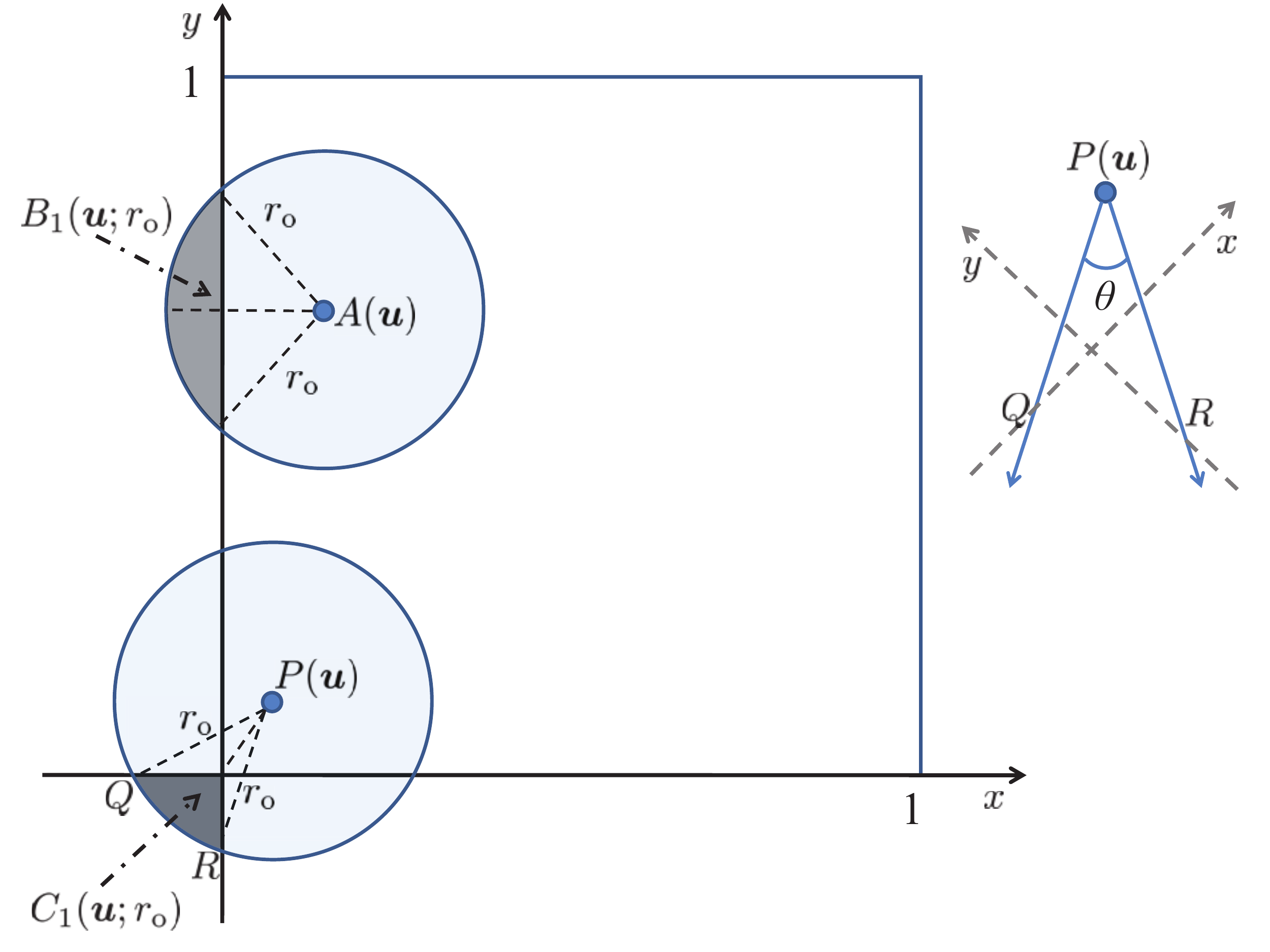}
    \caption{Illustration of border and corner effects in a unit square.}  \label{fig:one}
\end{figure*}
\else
\begin{figure}[t]
 \centering
    \includegraphics[width=0.45\textwidth]{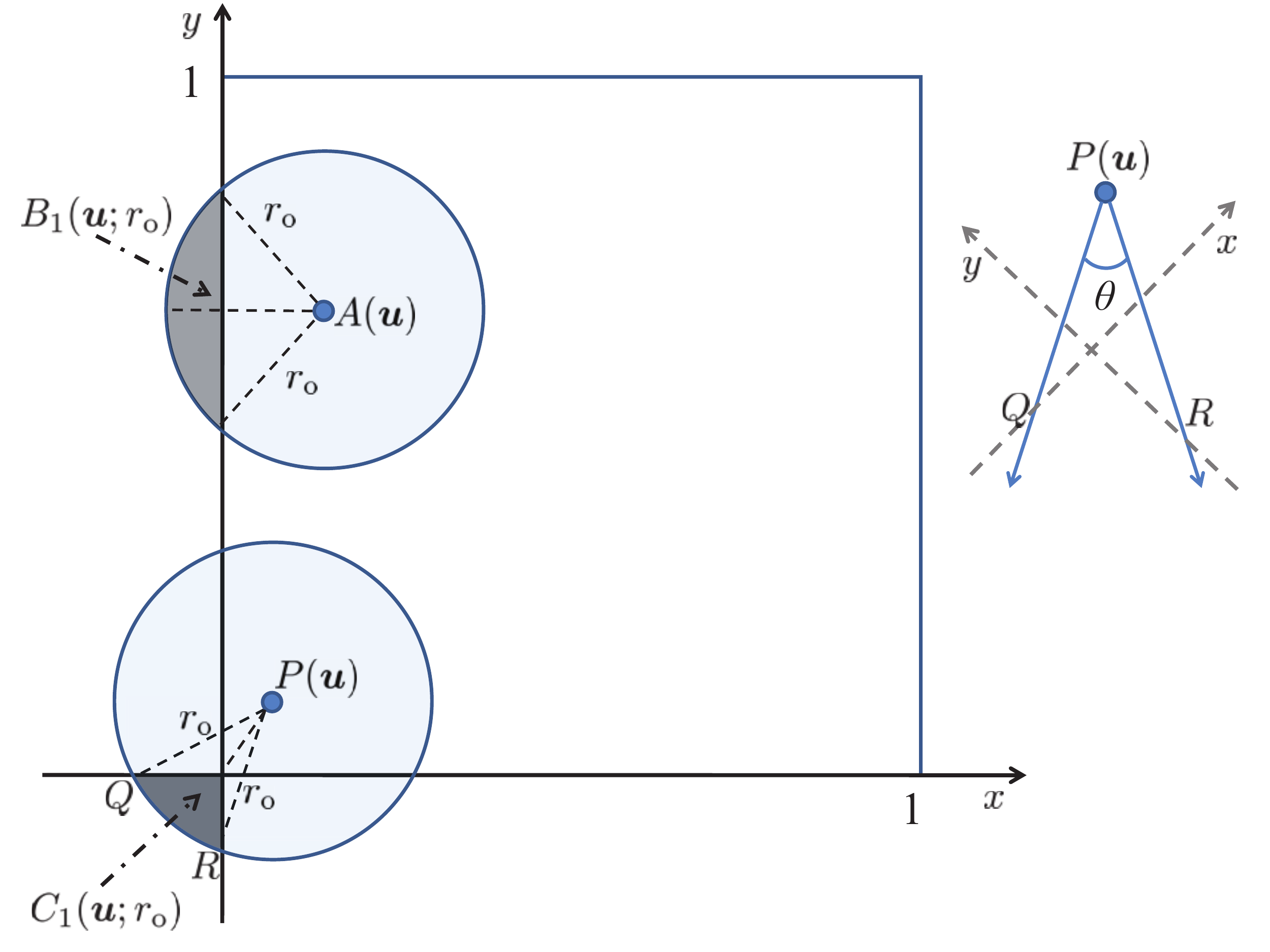}
    \caption{Illustration of border and corner effects in a unit square.}  \label{fig:one}
\end{figure}
\fi

\section{Proposed Framework}\label{sec:framework}

\subsection{Boundary Effects} We use the approach suggested in
\cite{Zubair-2012} in order to quantify the overlap area
$|\mathcal{O}(\dv{u};\rr)\cap\mathcal{R}|$. The basic building
blocks in this approach to characterize the boundary effects are (i)
the circular segment areas formed outside each side (\textit{border
effects}) and (ii) the corner overlap areas between two circular
segments formed at each vertex (\textit{corner effects}). We modify
the approach in~\cite{Zubair-2012} by placing the origin at the
vertex $V_1$, rather than at the center of the square. This leads to
a simpler formulation, as discussed below.

Let ${B}_{1}(\dv{u};\rr)$ denote the area of the circular segment
formed outside the side $S_{1}$, as illustrated in \figref{fig:one}.
Using the fact that the area of the circular segment is equal to the
area of the circular sector minus the area of the triangular
portion, we obtain
\begin{align}
\rba{1}\,= \,\begin{cases}
 \rr^2 \arccos \left(\frac{x}{\rr}\right)- \\ \quad x\sqrt{\rr^2-x^2} & \quad  \Delta_s(\dv{u},S_1) = x \geq \rr, \\
 0 & \textrm{otherwise},
 \label{bordereffects1}
\end{cases}
\end{align}

\noindent where $\Delta_s(\dv{u},S_\ell)$ denotes the Euclidean
distance between $\dv{u}$ and side $S_\ell,\,\ell=1,2,3,4$.
Similarly, the areas of the circular segments formed outside the
sides $S_{2}$, $S_{3}$ and $S_{4}$, respectively, can be expressed
as
\ifCLASSOPTIONonecolumn
\begin{align}
\rba{2}\,&=\, \begin{cases}\rr^2 \arccos\left(\frac{y}{\rr}\right) -
y\sqrt{\rr^2-y^2},& \quad  \Delta_s(\dv{u},S_2) = y \geq \rr \\
 0, & \textrm{otherwise}, \end{cases} \label{bordereffects2}
\\
\rba{3} \,&=\,\begin{cases}\rr^2 \arccos \left(\frac{1-x}{\rr}\right) -  (1-x)\sqrt{\rr^2-(1-x)^2},& \quad  \Delta_s(\dv{u},S_3) = 1-x \geq \rr \\
 0, & \textrm{otherwise}, \end{cases} \label{bordereffects3}
\\
\rba{4}\,&=\,\begin{cases} \rr^2 \arccos
\left(\frac{1-y}{\rr}\right)
-(1-y)\sqrt{\rr^2-(1-y)^2},& \quad  \Delta_s(\dv{u},S_4) = 1-y \geq \rr \\
 0, & \textrm{otherwise}. \end{cases}
\label{bordereffects4}
\end{align}
\else
\begin{align}
\rba{2}\,&=\,\rr^2 \arccos\left(\frac{y}{\rr}\right) -  y\sqrt{\rr^2-y^2}, \label{bordereffects2}\\
\rba{3} \,&=\,\rr^2 \arccos \left(\frac{1-x}{\rr}\right)  \nonumber \\
 & \, -(1-x)\sqrt{\rr^2-(1-x)^2}, \label{bordereffects3}\\
\rba{4}\,&=\,\rr^2 \arccos \left(\frac{1-y}{\rr}\right) \nonumber \\
 & \, -(1-y)\sqrt{\rr^2-(1-y)^2}\label{bordereffects4}.
\end{align}
\fi

Let ${C}_{1}(\dv{u};\rr)$ denote the area of the corner
overlap region between two circular segments at vertex $V_{1}$, as
illustrated in \figref{fig:one}. Using the fact that the area of the
overlap region is equal to the area of the circular sector minus the
area of two triangular portions, we can easily show that
\ifCLASSOPTIONonecolumn
\begin{align}
\rca{1}\,&=\, \frac{1}{2}\rr^2\theta - \frac{1}{2}\left(\sqrt{\rr^2-y^2}-x\right)y -\frac{1}{2}\left(\sqrt{\rr^2-x^2}-y\right)x, \label{cornereffects1} %
\end{align}
\else
\begin{align}
\rca{1}\,&=\, \frac{1}{2}\rr^2\theta - \frac{1}{2}\left(\sqrt{\rr^2-y^2}-x\right)y \nonumber \\
         & \: -\frac{1}{2}\left(\sqrt{\rr^2-x^2}-y\right)x, \label{cornereffects1} %
\end{align}
\fi

\noindent where the angle $\theta$ is given by
\begin{align}
\theta\,&=\,
2\arcsin\left(\frac{\mathrm{abs}\left(\sqrt{\theta_1}\,\right)}{2\rr}\right),
\end{align}

\noindent where $\mathrm{abs}(\cdot)$ denotes the absolute value or
modulus and $\theta_1 = 2\rr^2-2x\sqrt{\rr^2-y^2}-
2y\sqrt{\rr^2-x^2}$. Similarly, the areas of the corner overlap
region formed at vertex $V_{2}$, $V_{3}$ and $V_{4}$, respectively,
can be expressed as
\ifCLASSOPTIONonecolumn
\begin{align}
\rca{2}\,&=\, \frac{1}{2}\rr^2\alpha - \frac{1}{2}\left(\sqrt{\rr^2-y^2}-(1-x)\right)y -\frac{1}{2}\left(\sqrt{\rr^2-(1-x)^2}-y\right)(1-x),\label{cornereffects2}\\ %
\rca{3}\,&=\, \frac{1}{2}\rr^2\beta -
\frac{1}{2}\left(\sqrt{\rr^2-(1-y)^2}-(1-x)\right)(1-y) -
\frac{1}{2}\left(\sqrt{\rr^2-(1-x)^2}-(1-y)\right)(1-x),
\label{cornereffects3}\\
\rca{4}\,&=\, \frac{1}{2}\rr^2\gamma -
\frac{1}{2}\left(\sqrt{\rr^2-(1-y)^2}-x\right)(1-y) -
\frac{1}{2}\left(\sqrt{\rr^2-x^2}-(1-y)\right)x,
\label{cornereffects4}
\end{align}
\else
\begin{align}
\rca{2}\,&=\, \frac{1}{2}\rr^2\alpha - \frac{1}{2}\left(\sqrt{\rr^2-y^2}-(1-x)\right)y \nonumber \\
& \: -\frac{1}{2}\left(\sqrt{\rr^2-(1-x)^2}-y\right)(1-x),\label{cornereffects2}\\ %
\rca{3}\,&=\, \frac{1}{2}\rr^2\beta -
\frac{1}{2}\left(\sqrt{\rr^2-(1-y)^2}-(1-x)\right)(1-y) \nonumber \\
& \: -\frac{1}{2}\left(\sqrt{\rr^2-(1-x)^2}-(1-y)\right)(1-x),
\label{cornereffects3}\\
\rca{4}\,&=\, \frac{1}{2}\rr^2\gamma -
\frac{1}{2}\left(\sqrt{\rr^2-(1-y)^2}-x\right)(1-y) \nonumber \\
& \: -\frac{1}{2}\left(\sqrt{\rr^2-x^2}-(1-y)\right)x,
\label{cornereffects4}
\end{align}
\fi

\noindent where the angles $\alpha$, $\beta$ and $\gamma$ are given
by
\begin{align}
\alpha\,&=\,2\arcsin\left(\frac{\mathrm{abs}\left(\sqrt{\alpha_1}\right)}{2\rr} \right)\\
\beta\,&=\,2\arcsin\left(\frac{\mathrm{abs}\left(\sqrt{\beta_1}\right)}{2\rr}\right)\\
\gamma\,&=\,2\arcsin\left(\frac{\mathrm{abs}\left(\sqrt{\gamma_1}\right)}{2\rr}\right)
\end{align}

\noindent where $\alpha_1 = 2\rr^2-2(1-x)\sqrt{\rr^2-y^2}-
2y\sqrt{\rr^2-(1-x)^2}$, $\beta_1=2\rr^2-2(1-x)\sqrt{\rr^2-(1-y)^2}-
2(1-y)\sqrt{\rr^2-(1-x)^2}$ and
$\gamma_1=2\rr^2-2x\sqrt{\rr^2-(1-y)^2}- 2(1-y)\sqrt{\rr^2-x^2}$. We
note that the expressions for $\rca{\ell}$ are valid only when
$\Delta_s(\dv{u},V_\ell)\geq \rr$, where $\Delta_s(\dv{u},V_\ell)$
denotes the Euclidean distance between $\dv{u}$ and vertex
$V_\ell,\,\ell=1,2,3,4$.. For the case when
$\Delta_s(\dv{u},V_\ell)<\rr$, $\rca{\ell}=0$.

Using \eqref{bordereffects1}$-$\eqref{bordereffects4} and
\eqref{cornereffects1}$,~$\eqref{cornereffects2}$-$\eqref{cornereffects4},
the CDF $F(\dv{u};\rr)$ in \eqref{Eq:CDF_general} can be expressed
in closed-form, e.g., if $\rr=0.1$ and $\dv{u}=(0,0)$, then two
circular segments are formed outside sides $S_1$ and $S_2$ and also
there is overlap between them. Hence, in this case, $F(\dv{u};\rr) =
\pi \rr^2 -(B_1(\dv{u};\rr) + B_2(\dv{u};\rr) - C_1(\dv{u};\rr))$.
This will be further illustrated in the next subsection.

\subsection{Tractable Framework}

As illustrated in the last subsection, for a given value of the
transmission range $\rr$ and the location of the arbitrary node
$\dv{u}$, $F(\dv{u};\rr)$ can be expressed in closed-form using
\eqref{bordereffects1}$-$\eqref{bordereffects4} and
\eqref{cornereffects1}$,~$\eqref{cornereffects2}$-$\eqref{cornereffects4}.
In order to facilitate the averaging of \eqref{Eq:CDF_general} over
the whole square region, we divide the square region into different
non-overlapping subregions based on the different border and corner
effects that occur in that region. Due to the symmetry of the
square, some subregions have the same number of border and corner
effects which can be exploited to further simplify the averaging.
This will be elaborated in detail shortly.

Let $\mathcal{R}_1$, $\mathcal{R}_2$, $\hdots$ $\mathcal{R}_M$
denote the type of non-overlapping subregions and $n_i$,
$i\in\{1,\,2\,\hdots,\, M\}$ denote the number of subregion of type
$\mathcal{R}_i$. If $F_i(\dv{u};\rr)$ denotes the conditional
probability of connectivity for a node located at $\dv{u} \in
\mathcal{R}_i$, we can write the probability of node isolation in
\eqref{Eq:piso} as
\begin{align}\label{Eq:p_iso_general_sum}
\pis\,=\, \sum_{i=1}^{M} n_i\,\displaystyle\int_{\mathcal{R}_i}
\big( 1 - F_i(\dv{u};\rr) \big)^{N-1} ds(\dv{u}),
\end{align}
and the minimum node degree distribution $f_D(k;\rr)$ in
\eqref{Eq:min_node_degree} as
\begin{align}\label{Eq:min_node_degree_sum}
f_D(k; \rr) &= \textrm{P}(D=k) =  \bigg(1- \sum_{d=0}^{k-1}
\sum_{i=1}^{M} n_i \binom{N-1}{d}
\,\times \nonumber \\ &\displaystyle\int_{\mathcal{R}_i} \left( F(\dv{u};\rr) \right)^{d}
\left(1- F(\dv{u};\rr) \right)^{n-d-1} ds(\dv{u}) \bigg)^N.
\end{align}

We note that the average node degree denoted by $\overline{D}$ can
also be determined using our framework as~\cite{Bettstetter:2004}
\begin{align}\label{Eq:node_degree_mean}
\overline{D} &= (N-1)\int_{\mathcal{R}}F(\dv{u};\rr) ds(\dv{u}) \nonumber \\
&=
(N-1)\sum_{i=1}^{M} n_i\,\displaystyle\int_{\mathcal{R}_i}
F(\dv{u};\rr) ds(\dv{u}).
\end{align}

\noindent In fact $\int_{\mathcal{R}}F(\dv{u};\rr)ds(\dv{u})$ in
\eqref{Eq:node_degree_mean} denotes the cumulative distribution
function of the distance between two randomly placed nodes and the
closed form analytical results exist in the literature for square,
hexagon~\cite{Fan2007} and convex regular polygons~\cite{Basel-2012}.

\begin{remark}
The general formulation for $\pis$ in \eqref{Eq:piso} is also
indirectly suggested in~\cite{Eslami-2012}. However, no guidelines
are presented in order to evaluate \eqref{Eq:piso}. Hence, the
authors in~\cite{Eslami-2012} use a deterministic grid deployment of
nodes to approximate the boundary effects when nodes are uniformly
and independently distributed in a square region. By contrast, we
provide a tractable framework for complete and exact
characterization of the boundary effects in
\eqref{Eq:p_iso_general_sum}.
\end{remark}

\begin{remark}
While $F_i(\dv{u};\rr)$ in~\eqref{Eq:CDF_general} can be expressed
analytically in closed-form, the integration in
\eqref{Eq:p_iso_general_sum} and \eqref{Eq:min_node_degree_sum} does
not have a closed-form due to the $N-1$ factor in the exponent.
However, it can be easily evaluated numerically using the explicit
closed-form expressions for $F_i(\dv{u};\rr)$ for different
transmission ranges and different subregions. It must be noted that
numerical evaluation of two-fold integrals is widely practiced in
the literature~\cite{Simon-2005}.
\end{remark}

\begin{remark}
We have considered a unit square region for the sake of simplicity
in the proposed formulation. For the general case of a square of
side length $L$,
\eqref{Eq:p_iso_general_sum}-\eqref{Eq:node_degree_mean}
 can be used with appropriate scaling
of the transmission range as $\rr\rightarrow \rr/L$.
\end{remark}

Since the subregions are classified on the basis of the boundary
effects, the subregions change with the transmission range $\rr$. We
divide the range $\rr$ over the desired interval $0\leq\rr\leq
\sqrt{2}$, as explained in Section II-C, such that the boundary
effects are the same for the different subregions over each
subinterval of the transmission range. This is explained in detail
in the next section.

\section{Effect of Boundaries for the Different Transmission Range Cases}

\subsection{Transmission Range - $0\leq \rr \leq 1/2$ }

\begin{figure}[t]
    \centering
    \captionof{figure}*{Subregions}
    \vspace{-9mm}
    \includegraphics[width=0.50\textwidth]{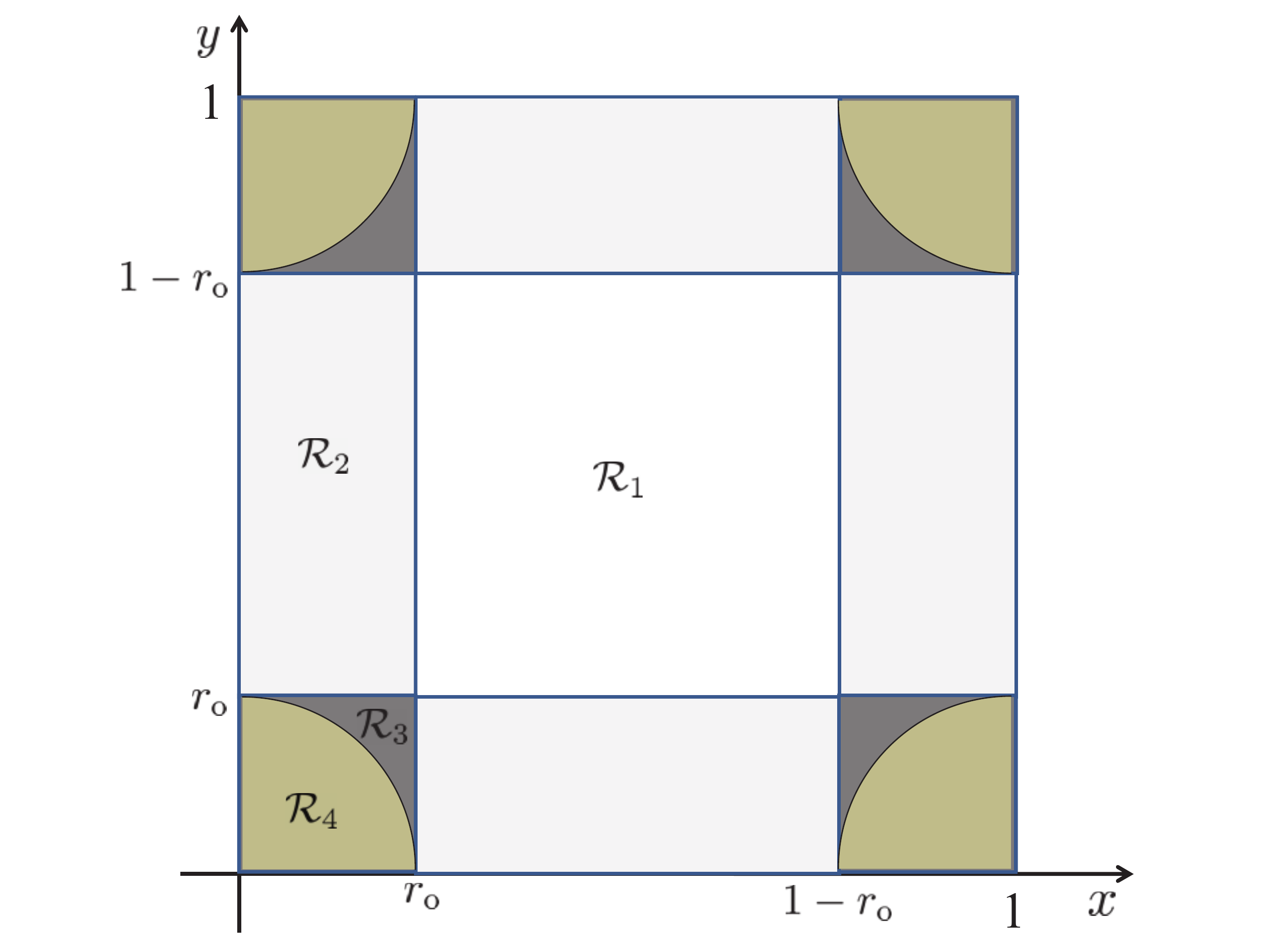}
    \small
    \label{table1}
      \captionof{table}{}
\begin{tabular}{l|m{0.2cm}|m{4cm}}
  \hline \hline
  $\mathcal{R}_i$  & $n_i$ & $F_i(\dv{u};\rr)$  \\ \hline \hline
 $\mathcal{R}_1$ &  $1$ & $\pi \rr^2$\\ \hline
 $\mathcal{R}_2$ &  $4$ & $\pi \rr^2-(B_1)$ \\ \hline
 $\mathcal{R}_3$ &  $4$ & $\pi \rr^2-(B_1+B_2)$ \\ \hline
 $\mathcal{R}_4$ &  $4$ & $\pi \rr^2-(B_1+B_2-C_1)$ \\ \hline \hline
\end{tabular}
\vspace{4mm}
\captionof{figure}{Subregions for transmission range $0 \leq \rr \leq
\frac{1}{2}$ are shown in the figure~(top) and conditional
probabilities $F_i(\dv{u};\rr)$ and number of subregions $n_i$ for each subregion are shown in the Table I (bottom).}
\label{fig:ranges_one}
    \end{figure}

Consider the first case of the transmission range, i.e., $0 \le \rr
\le \frac{1}{2}$, as shown in \figref{fig:ranges_one}. This case may
be of greatest interest in many practical situations where typically
the sensor transmission range is a small fraction of the side length
of the square. In this case, we can divide the square into
four~($M=4$) types of subregions $\mathcal{R}_1$, $\mathcal{R}_2$,
$\mathcal{R}_3$ and $\mathcal{R}_4$. As shown in
\figref{fig:ranges_one}, although there is one
subregion of type $\mathcal{R}_1$, there are four subregions of
types $\mathcal{R}_2$, $\mathcal{R}_3$ and $\mathcal{R}_4$,
respectively, which are shaded in the same color for ease of
identification, e.g., for an arbitrary node located in any subregion
of type $\mathcal{R}_2$, the disk $\mathcal{O}(\dv{u};\rr)$ is
limited by one side only. Hence, we determine $F_i(\dv{u};\rr)$
only for the following subregions
\begin{xitemize}
\item $\mathcal{R}_1 = \{ x\in(\rr,1-\rr) ,\,y\in(\rr,1-\rr)\}$

\item $\mathcal{R}_2 = \{ x\in(0,\rr) ,\,y\in(\rr,1-\rr)  \}$

\item $\mathcal{R}_3 = \{ x\in(0,\rr) ,\,y\in(\sqrt{\rr^2-x^2} ,\rr)\}$

\item $\mathcal{R}_4 = \{x\in( 0, \rr) ,\,y\in( 0,\sqrt{\rr^2-x^2})\}$
\end{xitemize}

It is easy to see that for an arbitrary node located anywhere in
subregion $\mathcal{R}_1$, the disk $\mathcal{O}(\dv{u};\rr)$ is
completely inside the square $\mathcal{R}$, i.e., there are no
border or corner effects. Hence, $F_1(\dv{u};\rr) = \pi \rr^2$. For
an arbitrary node located anywhere in subregion $\mathcal{R}_2$, the
disk $\mathcal{O}(\dv{u};\rr)$ is limited by side $S_1$, i.e., there
is a circular segment formed outside the side $S_1$. Hence,
$F_2(\dv{u};\rr) = \pi \rr^2 - (B_1(\dv{u};\rr))$. For an arbitrary
node located anywhere in subregion $\mathcal{R}_3$, the disk
$\mathcal{O}(\dv{u};\rr)$ is limited by sides $S_1$ and $S_2$, i.e.,
there is are two circular segments formed outside the sides $S_1$
and $S_2$ and there is no corner overlap between them. Hence,
$F_3(\dv{u};\rr) = \pi \rr^2 - (B_1(\dv{u};\rr)+B_2(\dv{u};\rr))$.
For an arbitrary node located anywhere in subregion $\mathcal{R}_4$,
the disk $\mathcal{O}(\dv{u};\rr)$ is limited by sides $S_1$ and
$S_2$ and vertex $V_1$, i.e., there is are two circular segments
formed outside the sides $S_1$ and $S_2$ and there is corner overlap
between them. Hence, $F_4(\dv{u};\rr) = \pi \rr^2 -
(B_1(\dv{u};\rr)+B_2(\dv{u};\rr)-C_1(\dv{u};\rr))$. The number of
subregions $n_i$ of each type and the corresponding closed form
$F_i(\dv{u};\rr)$ are tabulated in Table I. For the sake of brevity,
$B_\ell(u;r_0)$ and  $C_\ell(u;r_0)$ are denoted as  $B_\ell$ and
$C_\ell$, respectively in this and subsequent tables.

As $\rr$ increases from $0$ to ${1}/{2}$, we can see that the
subregions of type $\mathcal{R}_1$ and $\mathcal{R}_2$ become
smaller and the subregions of type $\mathcal{R}_3$ and
$\mathcal{R}_4$ become larger. For the value of range $\rr =
\frac{1}{2}$, the subregions of type $\mathcal{R}_1$ and
$\mathcal{R}_2$ approach zero.

\begin{remark}
The division of the square $\mathcal{R}$ into subregions for
transmission range $0\leq \rr \leq 1/2$ has been previously shown
in~\cite[Fig. 7]{Eslami-2012}, \cite[Fig. 2]{Mao-2012}
and~\cite[Fig. 2]{Nze-2011} to illustrate the intuitive
argument that the nodes situated in boundary subregions experience
border effects. However, subregions $\mathcal{R}_3$ and
$\mathcal{R}_4$ are indicated as one subregion
in~\cite{Mao-2012,Eslami-2012}. Using our framework, we show that
these are two distinct subregions with unique border and corner
effects. In addition, we formulate all the subregions for all
possible values of the range. This is different to
\cite{Nze-2011,Nze:2012} where only the transmission range $0 \leq
\rr \leq 1/2$ was considered for finding the average node degree.
\end{remark}


\subsection{Transmission Range - $1/2 \leq \rr \leq (2-\sqrt{2})$}

\begin{figure}[t]
    \centering
    \captionof{figure}*{Subregions}
    \vspace{-9mm}
    \includegraphics[width=0.50\textwidth]{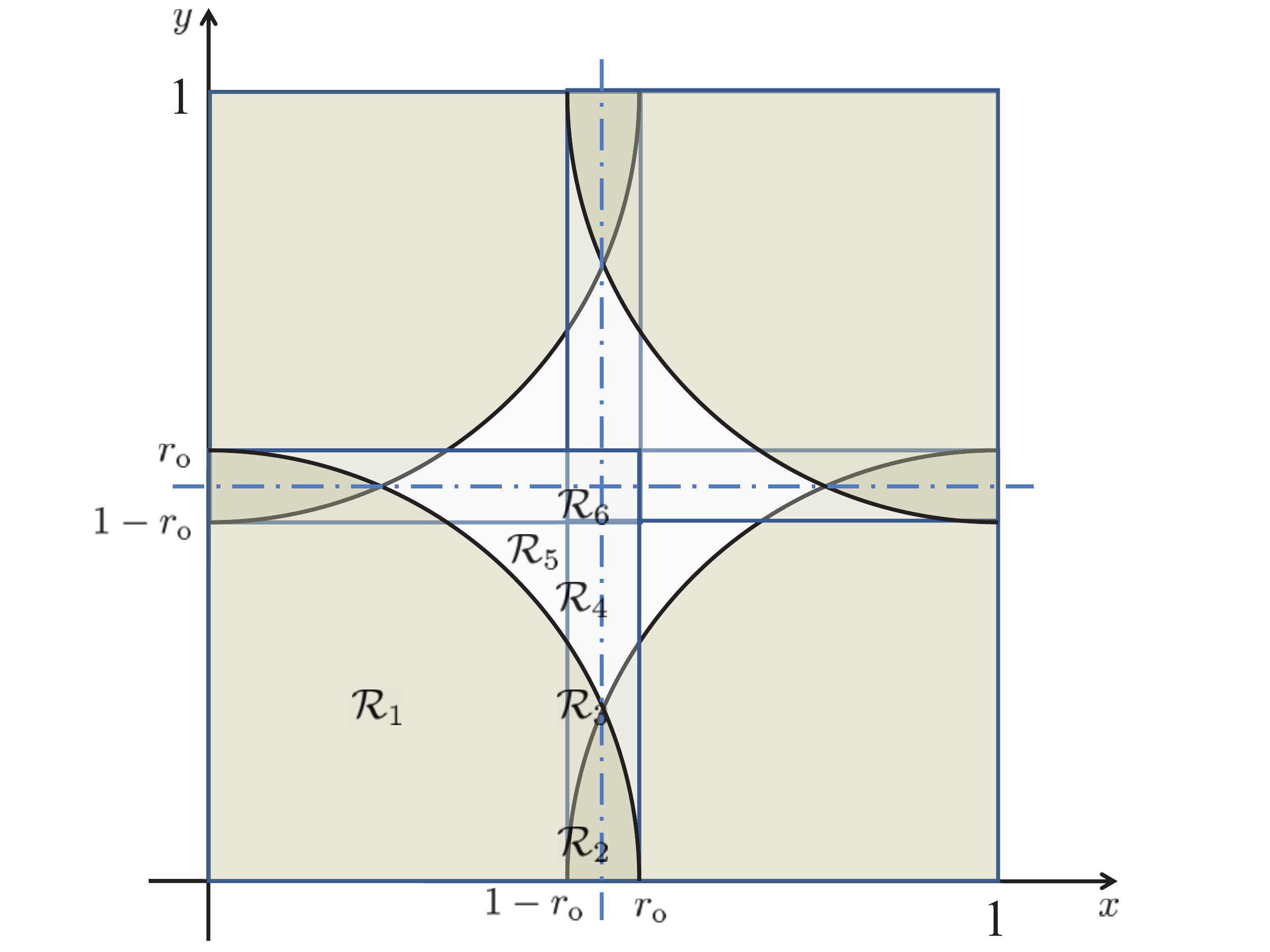}
    \small
    \label{table2}
      \captionof{table}{}
\begin{tabular}{l|m{0.2cm}|m{6cm}}
  \hline \hline
  $\mathcal{R}_i$  & $n_i$ & $F_i(\dv{u};\rr)$  \\ \hline \hline
    $\mathcal{R}_1$ & $4$ & $\pi \rr^2-(B_1+B_2-C_1)$\\ \hline
    $\mathcal{R}_2$ & $8$ & $\pi \rr^2-(B_1+B_2+B_3-C_1-C_2)$ \\ \hline
    $\mathcal{R}_3$ & $8$ & $\pi \rr^2-(B_1+B_2+B_3-C_1)$ \\ \hline
    $\mathcal{R}_4$ & $8$ & $\pi \rr^2-(B_1+B_2+B_3)$ \\ \hline
    $\mathcal{R}_5$ & $4$ & $\pi \rr^2-(B_1+B_2)$ \\ \hline
    $\mathcal{R}_6$ & $4$ & $\pi \rr^2-(B_1+B_2+B_3+B_4)$ \\ \hline\hline
\end{tabular}
\vspace{4mm}
\captionof{figure}{Subregions for transmission range $1/2 \leq \rr \leq (2-\sqrt{2})$ are shown in the figure~(top) and conditional probabilities $F_i(\dv{u};\rr)$ and number of subregions $n_i$ for each subregion are shown in the Table II (bottom).}
\label{fig:ranges_two}
    \end{figure}

For the case of the transmission range in the interval $1/2 \leq \rr
\leq 2-\sqrt{2}$, we have $M=6$ types of subregions, which are shown
in \figref{fig:ranges_two} and can be expressed as
\begin{xitemize}
\item $\mathcal{R}_1 = \{x\in(0,\sqrt{2\rr-1}) ,\,y\in(0,1-\rr)\}\cup \{x\in(\sqrt{2\rr-1},1-\rr),\,y\in(0,\sqrt{\rr^2-x^2}) \}$

\item $\mathcal{R}_2 = \{x\in(1-\rr,0.5) ,\,y\in(0,\sqrt{\rr^2-(x-1)^2})\} $

\item $\mathcal{R}_3 = \{x\in(1-\rr,0.5) ,\,y\in(\sqrt{\rr^2-(x-1)^2} ,\sqrt{\rr^2-x^2})\}$

\item $\mathcal{R}_4 = \{x\in(1-\rr,0.5) ,\,y\in( \sqrt{\rr^2-x^2},1-\rr)\}$

\item $\mathcal{R}_5 = \{x\in( \sqrt{2\rr-1},1- \rr) ,\,y\in( \sqrt{\rr^2-x^2},1-\rr)\} $

\item $\mathcal{R}_6 = \{x\in( 1-\rr,0.5) ,\,y\in( 1-\rr,0.5)\}$

\end{xitemize}

The upper limit for this interval of transmission range, i.e.,
$(2-\sqrt{2})$ is computed as the range $\rr$ for which the lines
$x=1-\rr$, $y=1-\rr$ and circle $x^2+y^2 = \rr^2$ intersect. As
$\rr$ approaches $(2-\sqrt{2})$, the subregion $\mathcal{R}_5$
squeezes to zero. The number of subregions $n_i$ of each type and
the corresponding closed-form $F_i(\dv{u};\rr)$ are tabulated in
Table~II.

\begin{figure}[t]
    \centering
    \captionof{figure}*{Subregions}
    \vspace{-9mm}
    \includegraphics[width=0.50\textwidth]{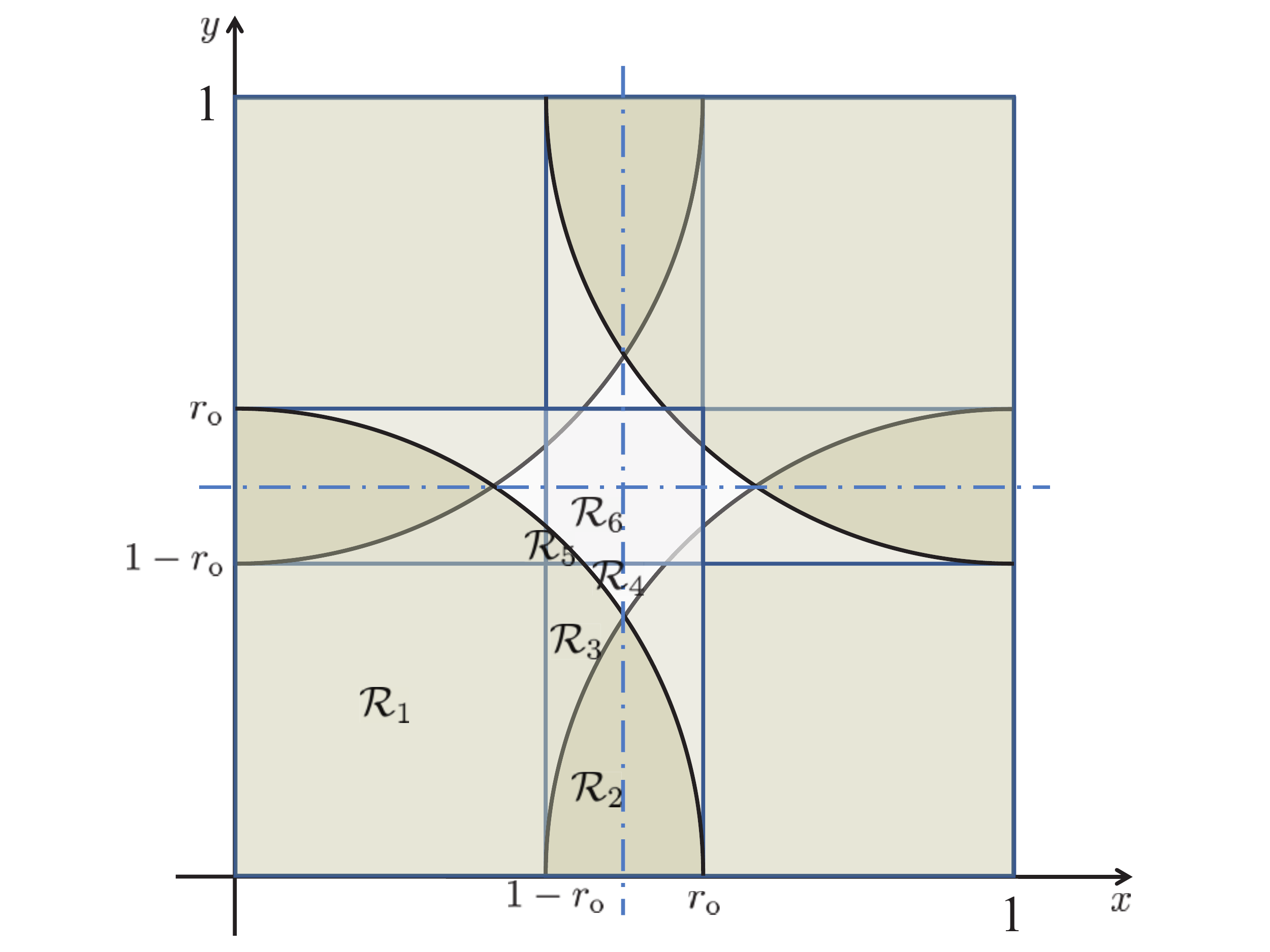}
    \small
    \label{table3}
      \captionof{table}{}
\begin{tabular}{l|m{0.2cm}|m{6cm}}
  \hline \hline
  $\mathcal{R}_i$  & $n_i$ & $F_i(\dv{u};\rr)$  \\ \hline \hline
    $\mathcal{R}_1$ & $4$ & $\pi \rr^2-(B_1+B_2-C_1)$\\ \hline
    $\mathcal{R}_2$ & $8$ & $\pi \rr^2-(B_1+B_2+B_3-C_1-C_2)$ \\  \hline
    $\mathcal{R}_3$ & $8$ & $\pi \rr^2-(B_1+B_2+B_3-C_1)$ \\ \hline
    $\mathcal{R}_4$ & $8$ & $\pi \rr^2-(B_1+B_2+B_3)$ \\ \hline
    $\mathcal{R}_5$ & $4$ & $\pi \rr^2-(B_1+B_2+B_3+B_4-C_1)$ \\ \hline
    $\mathcal{R}_6$ & $4$ & $\pi \rr^2-(B_1+B_2+B_3+B_4)$ \\ \hline\hline
\end{tabular}
\vspace{4mm}
\captionof{figure}{Subregions for transmission range $(2-\sqrt{2})\leq \rr \leq 5/8$ are shown in the figure~(top) and conditional probabilities $F_i(\dv{u};\rr)$ and number of subregions $n_i$ for each subregion are shown in the Table III (bottom).}
\label{fig:ranges_three}
    \end{figure}


\subsection{Transmission Range - $(2-\sqrt{2})\leq \rr \leq 5/8$ }

For the case of the transmission range in the interval
$(2-\sqrt{2})\leq \rr \leq 5/8$, we again have $M=6$ types of
subregions, which are shown in \figref{fig:ranges_three} and can be
expressed as

\begin{xitemize}

\item $\mathcal{R}_1 =  \{x\in(0,1-\rr) ,\,y\in(0,1-\rr)\} $

\item $\mathcal{R}_2 = \{x\in(1-\rr,0.5) ,\,y\in(0,\sqrt{\rr^2-(x-1)^2})\} $

\item $\mathcal{R}_3 = \{x\in(1-\rr,\sqrt{2\rr-1}) ,\,y\in(\sqrt{\rr^2-(x-1)^2} ,1-\rr)\}\cup\{x\in(\sqrt{2\rr-1},0.5) ,\,y\in(\sqrt{\rr^2-(x-1)^2} ,\sqrt{\rr^2-x^2})\}$

\item $\mathcal{R}_4 = \{x\in(\sqrt{2\rr-1},0.5) ,\,y\in(\sqrt{\rr^2-x^2},1-\rr)\}$

\item $\mathcal{R}_5 = \{x\in( 1- \rr, \sqrt{2\rr-1}) ,\,y\in( 1-\rr,\sqrt{\rr^2-x^2})\}$

\item $\mathcal{R}_6 = \{x\in( 1-\rr,\sqrt{2\rr-1}) ,\,y\in( \sqrt{\rr^2-x^2},0.5)\}\cup \{x\in( \sqrt{2\rr-1},0.5) ,\,y\in( 1-\rr,0.5)\}$
\end{xitemize}

The upper limit for this interval of the transmission range, i.e., $5/8$
is determined as the range $\rr$ where the subregion $\mathcal{R}_4$
squeezes to zero and is computed as an intersection of the line
$y=1-\rr$ and two circles $x^2+y^2 = \rr^2$ and $(x-1)^2+y^2 =
\rr^2$. The number of subregions $n_i$ of each type and the
corresponding closed-form $F_i(\dv{u};\rr)$ are tabulated in
Table~III.

\begin{figure}[t]
    \centering
    \captionof{figure}*{Subregions}
    \vspace{-9mm}
    \includegraphics[width=0.50\textwidth]{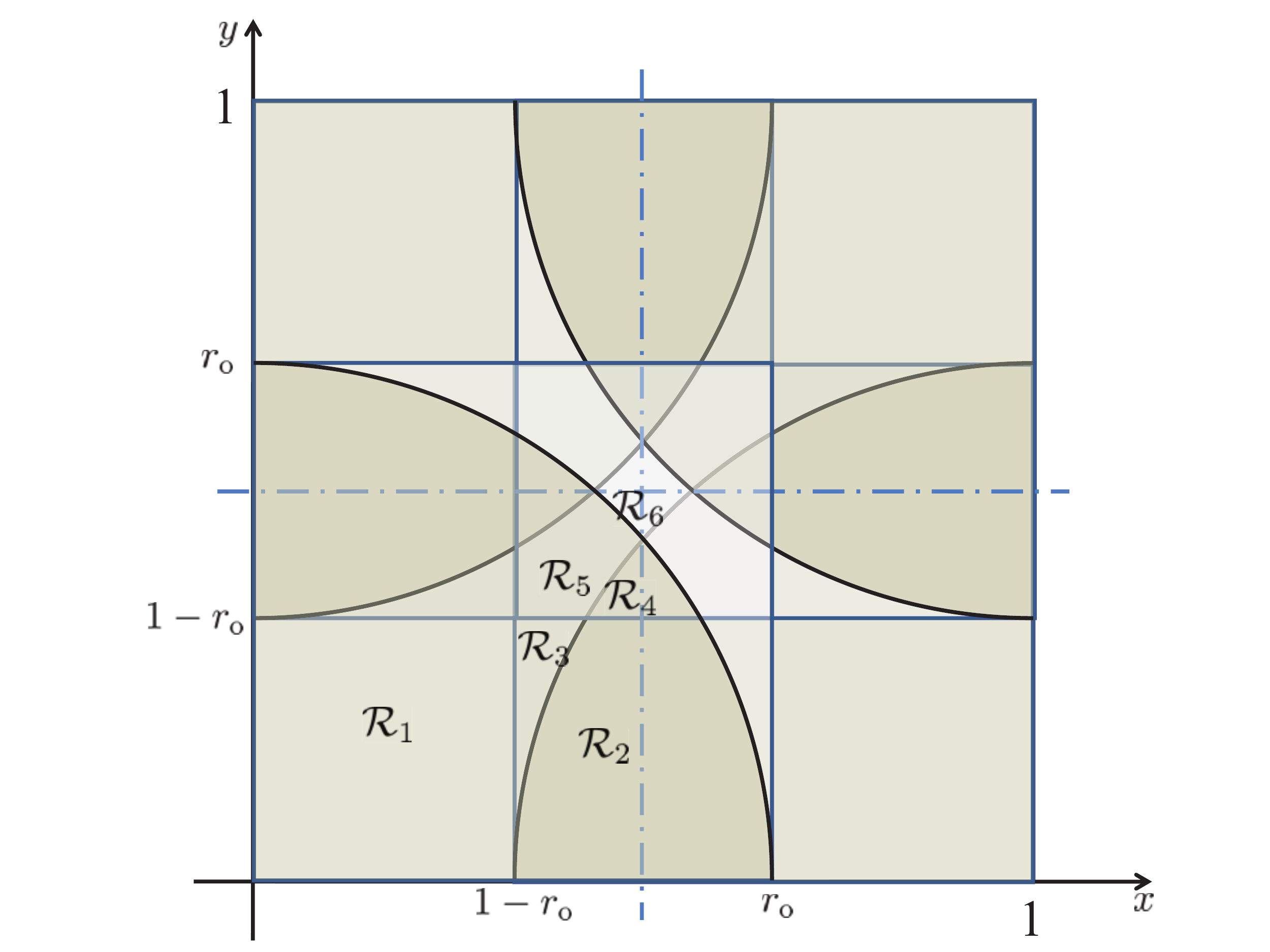}
    \small
    \label{table4}
      \captionof{table}{}
\begin{tabular}{l|m{0.2cm}|m{6cm}}
  \hline \hline
  $\mathcal{R}_i$  & $n_i$ & $F_i(\dv{u};\rr)$  \\ \hline \hline
 $\mathcal{R}_1$ & $4$ & $\pi \rr^2-(B_1+B_2-C_1)$\\ \hline
 $\mathcal{R}_2$ & $8$ & $\pi \rr^2-(B_1+B_2+B_3-C_1-C_2)$ \\ \hline
 $\mathcal{R}_3$ & $8$ & $\pi \rr^2-(B_1+B_2+B_3-C_1)$ \\ \hline
 $\mathcal{R}_4$ & $8$ & $\pi \rr^2-(B_1+B_2+B_3+B_4-C_1-C_2)$ \\ \hline
 $\mathcal{R}_5$ & $4$ & $\pi \rr^2-(B_1+B_2+B_3+B_4-C_1)$ \\ \hline
 $\mathcal{R}_6$ & $4$ & $\pi \rr^2-(B_1+B_2+B_3+B_4)$ \\ \hline\hline
\end{tabular}
\vspace{4mm}
\captionof{figure}{Subregions for transmission range $5/8\leq \rr \leq {1}/{\sqrt{2}}$ are shown in the figure~(top) and conditional probabilities $F_i(\dv{u};\rr)$ and number of subregions $n_i$ for each subregion are shown in the Table IV (bottom).}
\label{fig:ranges_four}
    \end{figure}

\subsection{Transmission Range - $5/8\leq \rr \leq {1}/{\sqrt{2}}$ }

For the case of the transmission range in the interval $5/8\leq \rr
\leq {1}/{\sqrt{2}}$, we have $M=6$ types of subregions, which
are shown in \figref{fig:ranges_four} and can be expressed as

\begin{xitemize}
\item $\mathcal{R}_1 = \{x\in(0,1-\rr) ,\,y\in(0,1-\rr)\} $

\item $\mathcal{R}_2 = \{x\in(1-\sqrt{\rr^2-y^2},0.5) ,\,y\in(0,1-\rr)\}$

\item $\mathcal{R}_3 = \{x\in(1-\rr,1-\sqrt{\rr^2-y^2}) ,\,y\in(0 ,1-\rr)\}$

\item $\mathcal{R}_4 = \{x\in(1-\sqrt{2\rr-1},0.5) ,\,y\in(1-\rr,\sqrt{\rr^2-(x-1)^2})\}$

\item $\mathcal{R}_5 = \{x\in( 1-\rr,\sqrt{\rr^2-0.25}) ,\,y\in(1-\rr,1- \sqrt{\rr^2-x^2})\}\cup\{x\in(\sqrt{\rr^2-0.25},1-\sqrt{2\rr-1}) ,\,y\in(1-\rr,\sqrt{\rr^2-x^2}) \}\cup\{x\in( 1-\sqrt{2\rr-1},0.5) ,\,y\in(\sqrt{\rr^2-(x-1)^2},\sqrt{\rr^2-x^2}) \} $

\item $\mathcal{R}_6 = \{x\in( \sqrt{\rr^2-0.25},0.5) ,\,y\in( \sqrt{\rr^2-x^2},0.5)\} $

\end{xitemize}

The upper limit for this interval of the transmission range, i.e.,
${1}/{\sqrt{2}}$ is determined as the range $\rr$ where the four
circles $x^2+y^2 = \rr^2$, $(x-1)^2+y^2 = \rr^2$, $(x-1)^2+(y-1)^2 =
\rr^2$ and $(x)2+(y-1)^2 = \rr^2$ intersect. The number of
subregions $n_i$ of each type and the corresponding closed-form
$F_i(\dv{u};\rr)$ are tabulated in Table~IV.

\begin{figure}[t]
    \centering
    \captionof{figure}*{Subregions}
    \vspace{-9mm}
    \includegraphics[width=0.50\textwidth]{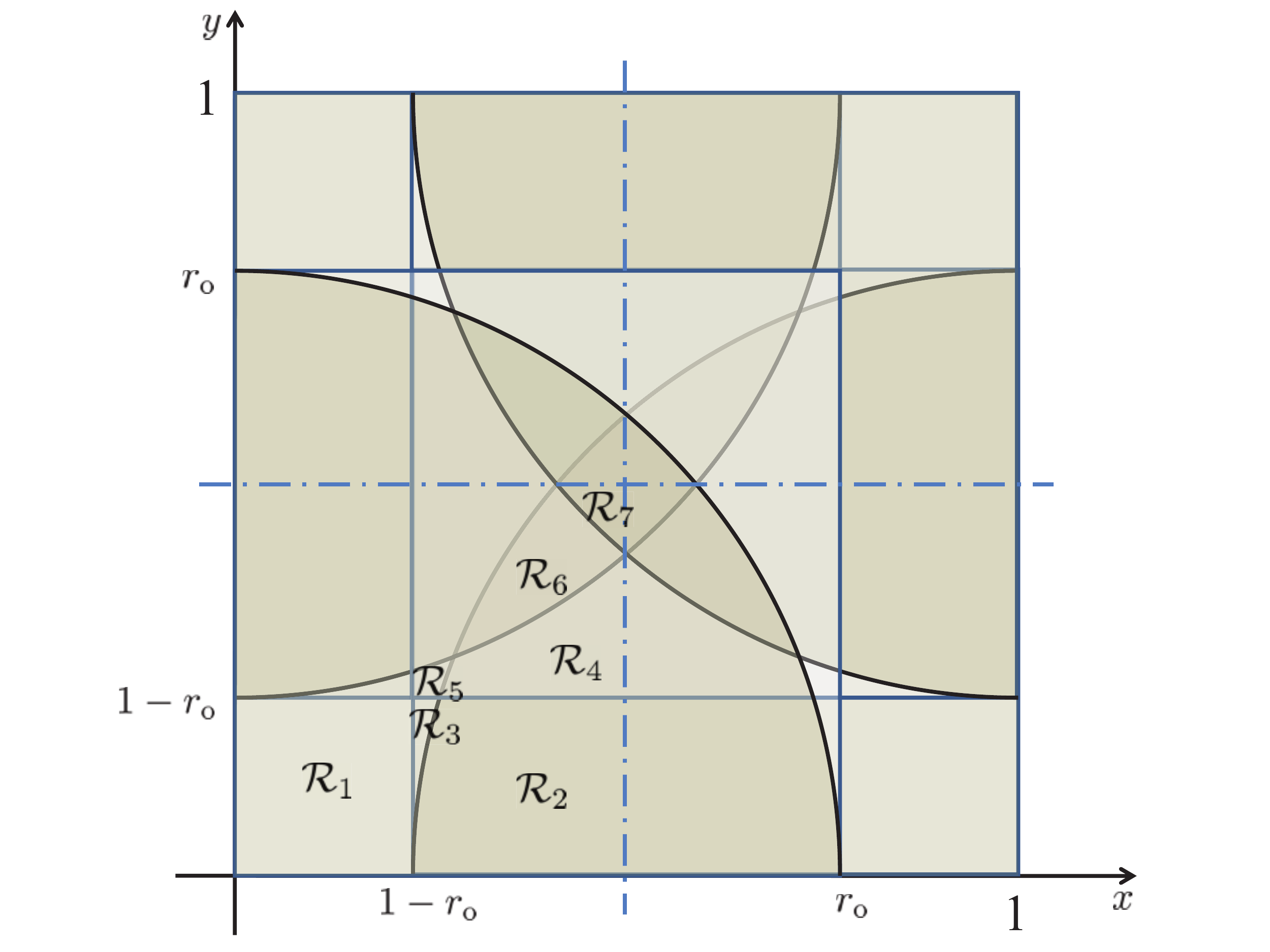}
    \small
    \label{table5}
      \captionof{table}{}
\begin{tabular}{l|m{0.2cm}|m{7cm}}
  \hline \hline
  $\mathcal{R}_i$  & $n_i$ & $F_i(\dv{u};\rr)$  \\ \hline \hline
 $\mathcal{R}_1$ & $4$ & $\pi \rr^2-(B_1+B_2-C_1)$\\ \hline
 $\mathcal{R}_2$ & $8$ & $\pi \rr^2-(B_1+B_2+B_3-C_1-C_2)$ \\ \hline
 $\mathcal{R}_3$ & $8$ & $\pi \rr^2-(B_1+B_2+B_3-C_1)$ \\ \hline
 $\mathcal{R}_4$ & $8$ & $\pi \rr^2-(B_1+B_2+B_3+B_4-C_1-C_2)$ \\ \hline
 $\mathcal{R}_5$ & $4$ & $\pi \rr^2-(B_1+B_2+B_3+B_4-C_1)$ \\ \hline
 $\mathcal{R}_6$ & $4$ & $\pi \rr^2-(B_1+B_2+B_3+B_4-C_1-C_2-C_4)$ \\ \hline
 $\mathcal{R}_7$ & $4$ & $\pi \rr^2-(B_1+B_2+B_3+B_4-C_1-C_2-C_3-C_4)=1$ \\ \hline\hline
\end{tabular}
\vspace{4mm}
\captionof{figure}{Subregions for transmission range $1/\sqrt{2}\leq \rr \leq 1$ are shown in the figure~(top) and conditional probabilities $F_i(\dv{u};\rr)$ and number of subregions $n_i$ for each subregion are shown in the Table V (bottom).}
\label{fig:ranges_five}
    \end{figure}

\subsection{Transmission Range - ${1}/{\sqrt{2}}\leq \rr \leq 1$ }

For the case of the transmission range in the interval
$1/\sqrt{2}\leq \rr \leq 1$, we have $M=7$ types of subregions,
which are shown in \figref{fig:ranges_five} and can be expressed as

\begin{xitemize}
\item $\mathcal{R}_1 =  \{x\in(0,1-\rr) ,\,y\in(0,1-\rr)\}$

\item $\mathcal{R}_2 = \{x\in(1-\rr,1-\sqrt{2\rr-1}) ,\,y\in(0,\sqrt{\rr^2-(x-1)^2})\}\cup\{x\in(1-\sqrt{2\rr-1},0.5) ,\,y\in(0,1-\rr)\}$

\item $\mathcal{R}_3 = \{x\in(1-\rr,1-\sqrt{2\rr-1}) ,\,y\in(\sqrt{\rr^2-(x-1)^2},1-\rr)\}$

\item $\mathcal{R}_4 = \{x\in(1-\sqrt{2\rr-1},(1-\sqrt{2\rr^2-1})/2) ,\,y\in(1-\rr,\sqrt{\rr^2-(x-1)^2} )\}\cup\{x\in((1-\sqrt{2\rr^2-1})/2,0.5) ,\,y\in( 1-\rr,1-\sqrt{\rr^2-x^2})\} $

\item $\mathcal{R}_5 = \{x\in(1-\rr,1-\sqrt{2\rr-1}) ,\,y\in(1-\rr,1-\sqrt{\rr^2-x^2})\}\cup\{x\in(1-\sqrt{2\rr-1},(1-\sqrt{2\rr^2-1})/2) ,\,y\in(\sqrt{\rr^2-(x-1)^2},1-\sqrt{\rr^2-x^2})\} $

\item $\mathcal{R}_6 = \{x\in((1-\sqrt{2\rr^2-1})/2,1-\sqrt{\rr^2-0.25}) ,\,y\in(1- \sqrt{\rr^2-x^2},\sqrt{\rr^2-(x-1)^2})\}\cup\{ x\in(1-\sqrt{\rr^2-0.25},0.5), y\in(1- \sqrt{\rr^2-x^2},1- \sqrt{\rr^2-(x-1)^2}) \} $

\item $\mathcal{R}_7 = \{x\in(1- \sqrt{\rr^2-0.25},0.5) ,\,y\in( 1-\sqrt{\rr^2-(x-1)^2},0.5)\}$
\end{xitemize}

The upper limit for this interval of the transmission range, i.e., $1$
corresponds to the length of the side of the square region. For
$\rr\geq 1$, there is always the effect of the sides of the square
on the coverage area of a node irrespective of the location of the node.
The number of subregions $n_i$ of each type and the corresponding
closed-form $F_i(\dv{u};\rr)$ are tabulated in Table~V.

\subsection{Transmission Range - $1\leq \rr \leq \sqrt{5}/2$ }

\begin{figure}[t]
    \centering
    \captionof{figure}*{Subregions}
    \vspace{-9mm}
    \includegraphics[width=0.50\textwidth]{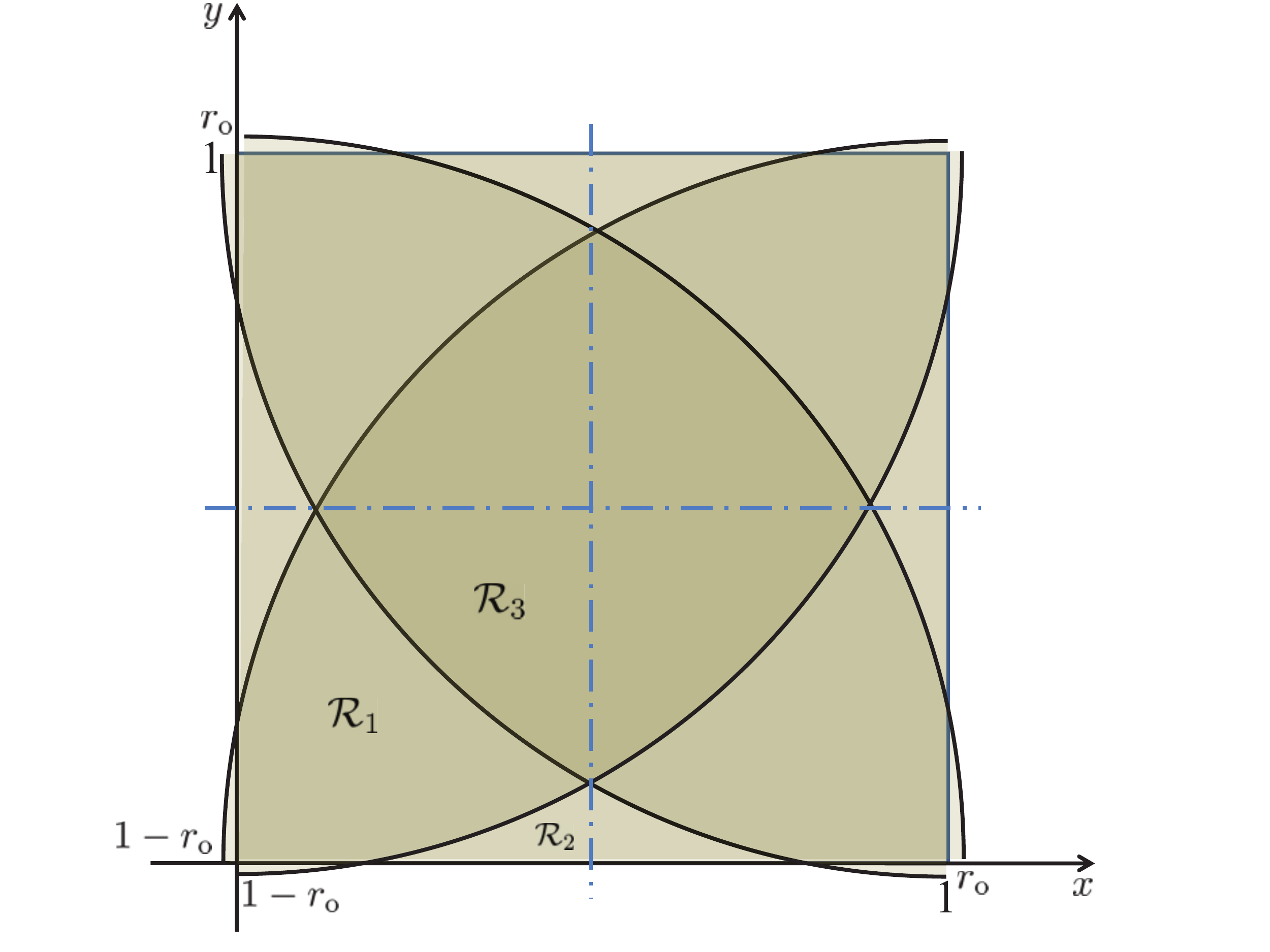}
    \small
    \label{table6}
      \captionof{table}{}
\begin{tabular}{l|m{0.2cm}|m{7cm}}
  \hline \hline
  $\mathcal{R}_i$  & $n_i$ & $F_i(\dv{u};\rr)$  \\ \hline \hline
 $\mathcal{R}_1$ & $4$ & $\pi \rr^2-(B_1+B_2+B_3+B_4-C_1-C_2-C_4)$\\ \hline
 $\mathcal{R}_2$ & $8$ & $\pi \rr^2-(B_1+B_2+B_3+B_4-C_1-C_2)$ \\ \hline
 $\mathcal{R}_3$ & $4$ & $\pi \rr^2-(B_1+B_2+B_3+B_4-C_1-C_2-C_3-C_4)=1$ \\ \hline \hline
\end{tabular}
\vspace{4mm}
\captionof{figure}{Subregions for transmission range $1 \leq \rr \leq \sqrt{5}/2$ are shown in the figure~(top) and conditional probabilities $F_i(\dv{u};\rr)$ and number of subregions $n_i$ for each subregion are shown in the Table VI (bottom).}
\label{fig:ranges_six}
    \end{figure}

For the case of the transmission range in the interval  $1\leq \rr
\leq \sqrt{5}/2$, we have $M=3$ types of subregions, which are shown
in \figref{fig:ranges_six} and can be expressed as

\begin{xitemize}
\item $\mathcal{R}_1 = \{x\in(0,1-\sqrt{\rr^2-0.25}) ,\,y\in(0,\sqrt{\rr^2-(x-1)^2})\}\cup\{x\in(1-\sqrt{\rr^2-0.25},\sqrt{\rr^2-1}) ,\,y\in(0,1-\sqrt{\rr^2-(x-1)^2}) \}\cup\{x\in(\sqrt{\rr^2-1},0.5),\,y\in(1- \sqrt{\rr^2-x^2}, 1- \sqrt{\rr^2-(x-1)^2})  \}$

\item $\mathcal{R}_2 = \{x\in(\sqrt{\rr^2-1},0.5) ,\,y\in(0,1- \sqrt{\rr^2-x^2})\} $

\item $\mathcal{R}_3 = \{x\in(1- \sqrt{\rr^2-0.25},0.5) ,\,y\in( 1-\sqrt{\rr^2-(x-1)^2},0.5)\}$

\end{xitemize}

The upper limit for this interval of the transmission range, i.e.,
$\sqrt{5}/2$ is determined as $\rr$ for which the circles
$(x-1)^2+(y-1)^2 = \rr^2$, $x^2+(y-1)^2 = \rr^2$ intersect and the
subregion $\mathcal{R}_2$ vanishes. The number of subregions $n_i$
of each type and the corresponding closed-form $F_i(\dv{u};\rr)$ are
tabulated in Table~VI.

%

\subsection{Transmission Range - $\sqrt{5}/2\leq \rr \leq \sqrt{2}$}

\begin{figure}[t]
    \centering
    \captionof{figure}*{Subregions}
    \vspace{-9mm}
    \includegraphics[width=0.50\textwidth]{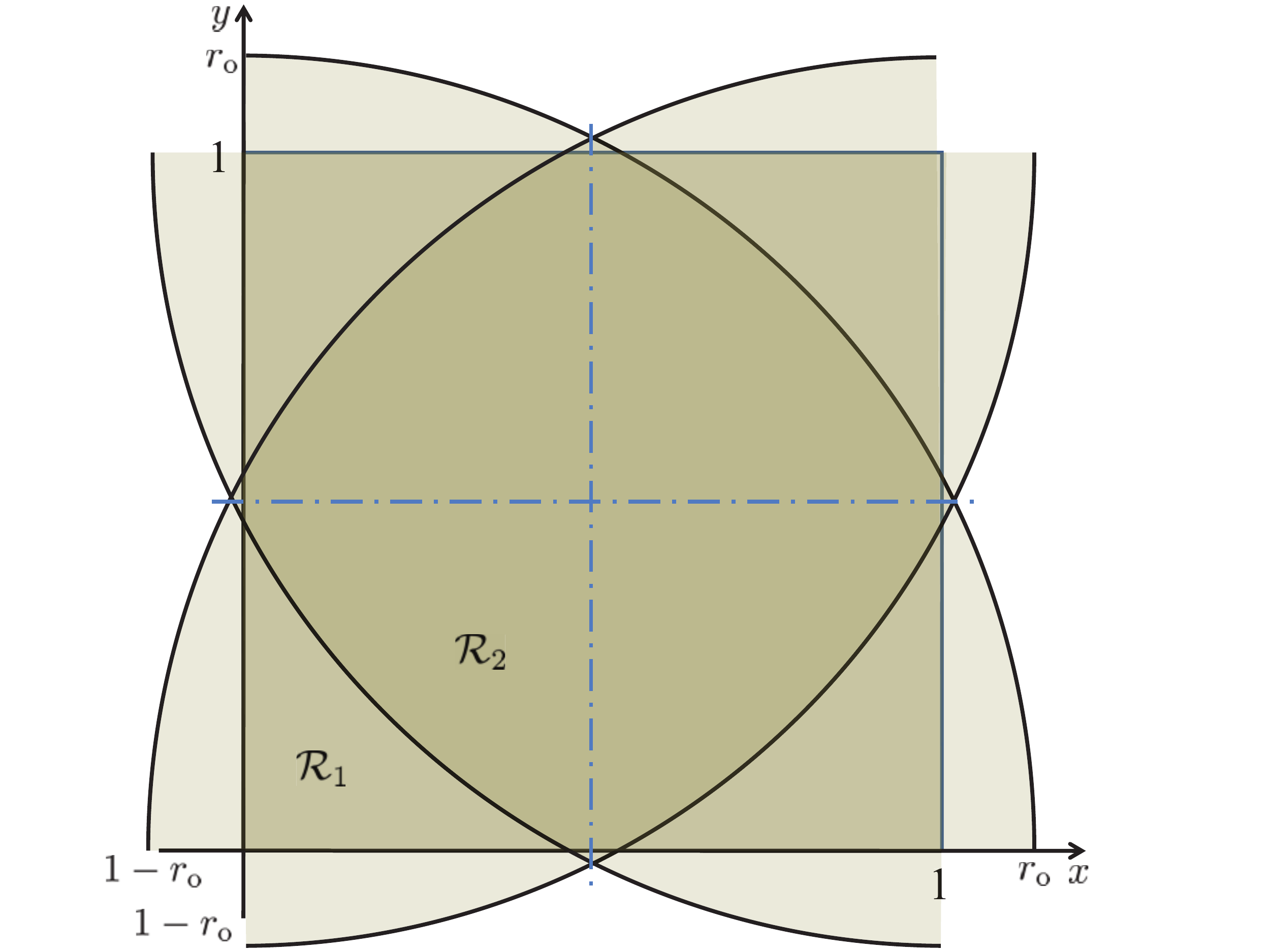}
    \small
    \label{table7}
      \captionof{table}{}
\begin{tabular}{l|m{0.2cm}|m{7cm}}
  \hline \hline
  $\mathcal{R}_i$  & $n_i$ & $F_i(\dv{u};\rr)$  \\ \hline \hline
 $\mathcal{R}_1$ & $4$ & $\pi \rr^2-(B_1+B_2+B_3+B_4-C_1-C_2-C_4)$\\ \hline
 $\mathcal{R}_2$ & $4$ & $\pi \rr^2-(B_1+B_2+B_3+B_4-C_1-C_2-C_3-C_4)=1$ \\ \hline\hline
\end{tabular}
\vspace{4mm}
\captionof{figure}{Subregions for transmission range $\sqrt{5}/2\leq \rr \leq \sqrt{2}$ are shown in the figure~(top) and conditional probabilities $F_i(\dv{u};\rr)$ and number of subregions $n_i$ for each subregion are shown in the Table VII (bottom).}
\label{fig:ranges_seven}
    \end{figure}

Finally, we have $M=2$ types of subregions for the case of the
transmission range in the interval $\sqrt{5}/2\leq \rr \leq
\sqrt{2}$. The regions are shown in \figref{fig:ranges_seven} and
can be expressed as

\begin{xitemize}
\item $\mathcal{R}_1 = \{x\in(0,1-\sqrt{\rr^2-1}) ,\,y\in(0,1-\sqrt{\rr^2-(x-1)^2})\}$
\item $\mathcal{R}_2 = \{x\in(0,1-\sqrt{\rr^2-1}) ,\,y\in(1- \sqrt{\rr^2-(x-1)^2},0.5)\}\cup\{x\in(1-\sqrt{\rr^2-1},0.5),\,y\in(0,0.5) \}$
\end{xitemize}
The number of subregions $n_i$ of each type and the corresponding
closed form $F_i(\dv{u};\rr)$ are tabulated in Table~VII. As
highlighted earlier, we note that the $F(u:\rr)=1$ for transmission
range $\rr$ greater than equal to $\sqrt{2}$~(diagonal length of the
square).

\section{Results}\label{sec:results}

In this section, we present the numerical results and compare with
the simulation results to validate the proposed framework. We also
compare with the results from the prior work to demonstrate the
advantage of our proposed framework, especially for smaller number
of sensor nodes $N$. We have implemented
\eqref{Eq:p_iso_general_sum} and \eqref{Eq:min_node_degree_sum} in
Mathematica. We consider the nodes to be independently and uniformly
distributed in a square region of side length $L=1$. The simulation
results are obtained by averaging over $S=50,000$ Monte Carlo
simulation runs.

\subsection{Probability of Node Isolation}
\begin{figure}[!t]
    \centering
    \includegraphics[width=0.70\textwidth]{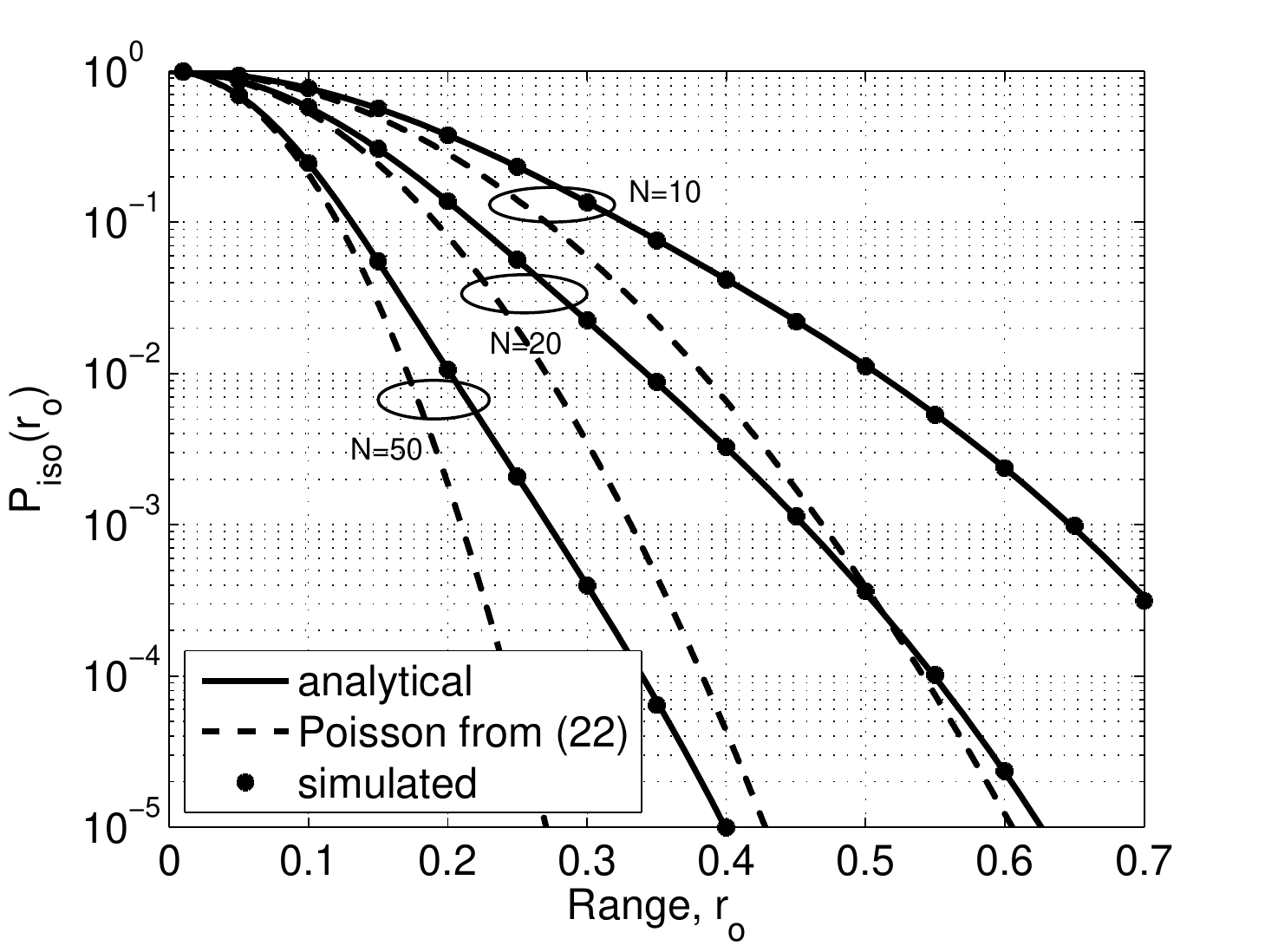}
     \caption{Probability of node isolation, $\pis$, versus the transmission range $\rr$ for $N=10,20,50$ nodes independently and uniformly distributed in a unit square.}
    \label{fig:nodes_iso}
\end{figure}

\figref{fig:nodes_iso} plots the probability of node isolation,
$\pis$, in~\eqref{Eq:p_iso_general_sum}, versus thetransmission range
$\rr$ for $N=10,20,50$ nodes. The probability of node isolation in
infinite homogenous Poisson point process
networks~\cite{bettstetter-2002}
\begin{align}
\pis &= e^{-\rho \pi \rr^2},
\end{align}

\noindent assuming constant node density $\rho = 10,20,50$
nodes/m$^2$ is also plotted as a reference. We can see from
\figref{fig:nodes_iso} that the simulation results for finite
networks match perfectly with the numerical results. This is to be
expected since we account for boundary effects accurately and
evaluate \eqref{Eq:p_iso_general_sum} exactly.
\figref{fig:nodes_iso} shows that the probability of node isolation
is greater in finite networks, compared with Poisson networks. This
is due to the inclusion of the border and corner effects, as
explained in Section~\ref{sec:framework}.

\subsection{Minimum Node Degree Distribution and $k$-Connectivity}\label{sec:results_min_node}

\begin{figure*}[t]
    \vspace{-4mm}
    \centering
    \subfloat[$N=10$]
        {\label{fig:node_degree_a}\includegraphics[scale=0.42]{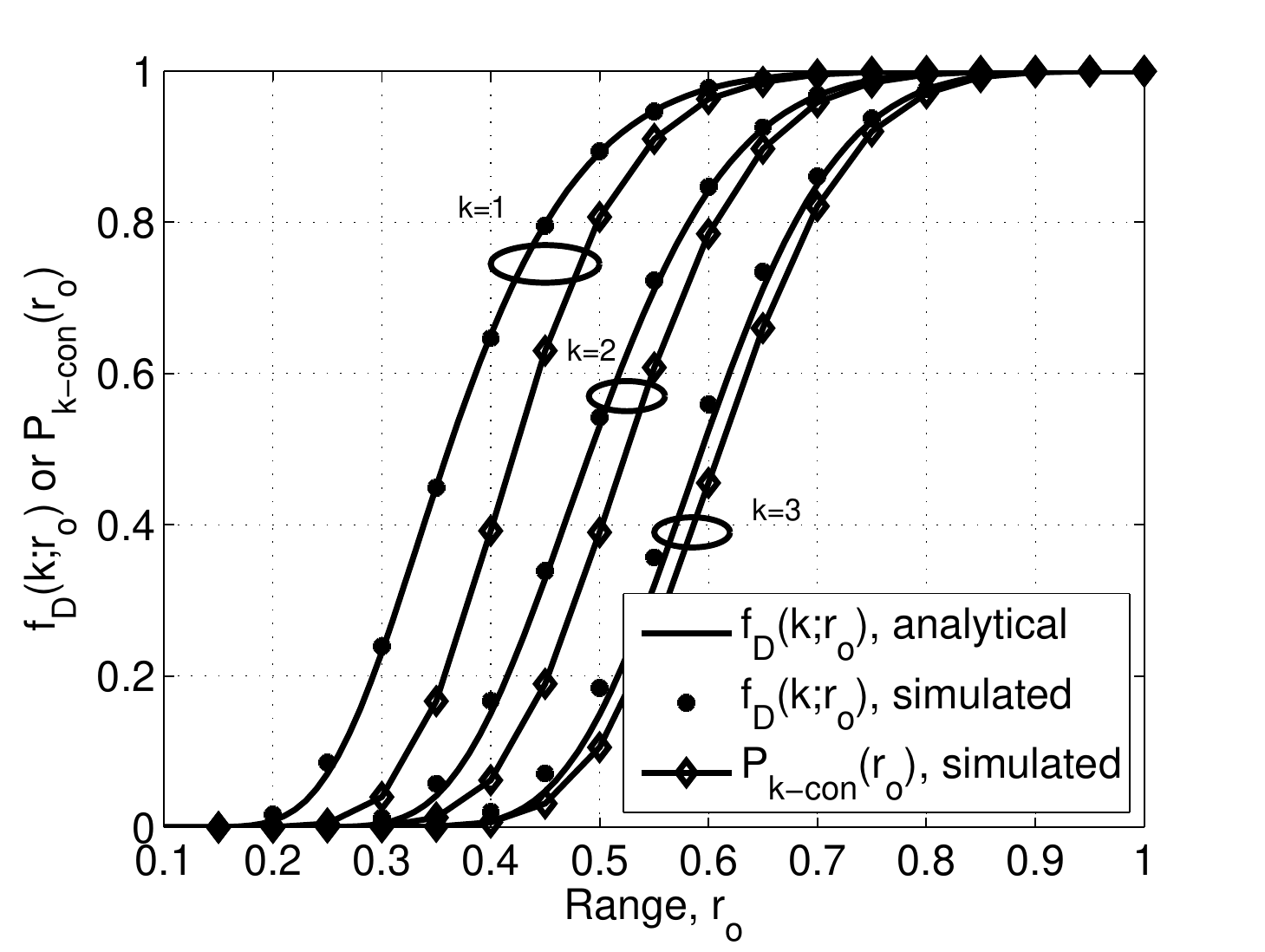}}
    \subfloat[$N=20$]
        {\label{fig:node_degree_b}\includegraphics[scale=0.42]{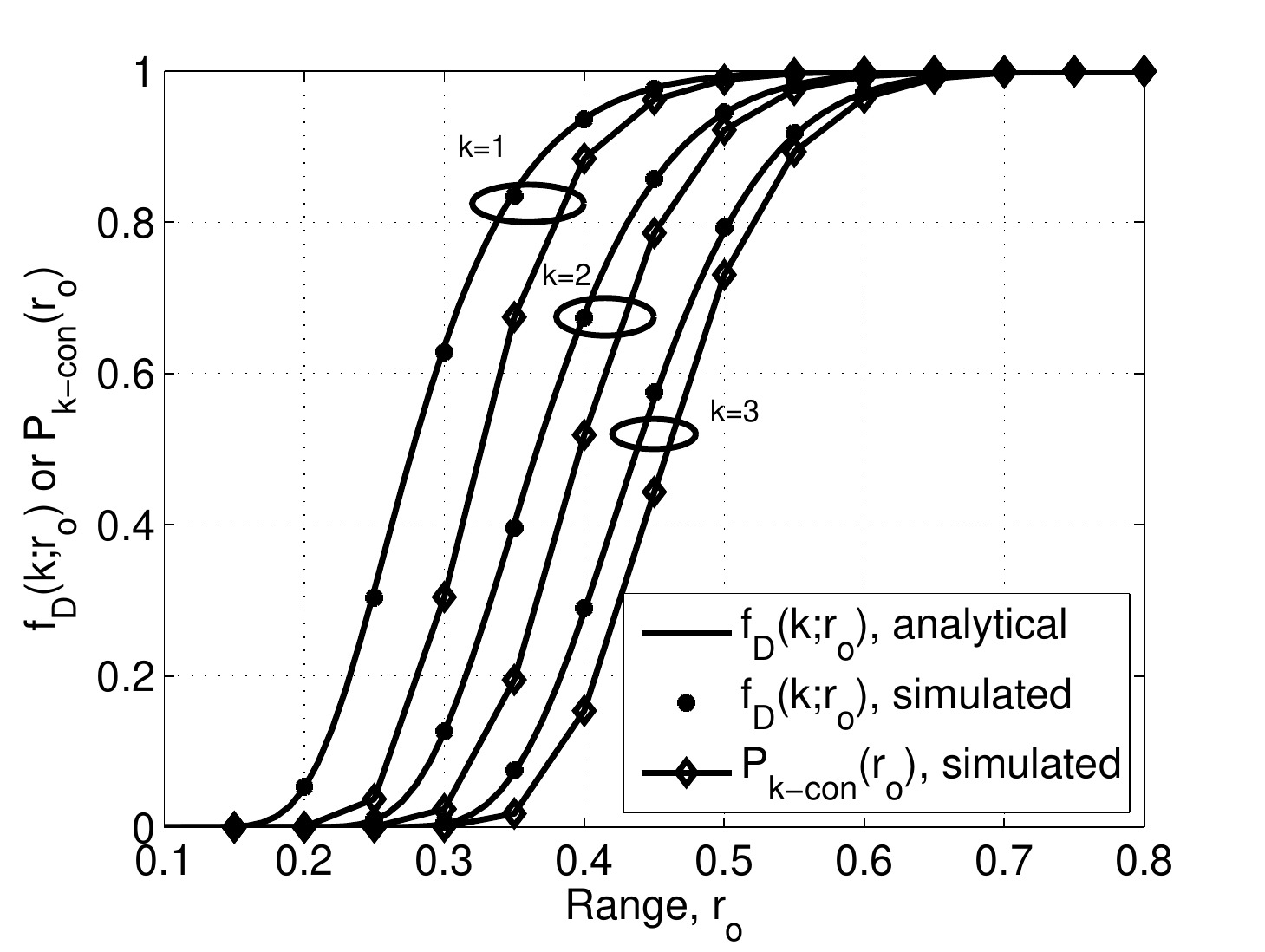}}
    \subfloat[$N=50$]
        {\label{fig:node_degree_c}\includegraphics[scale=0.42]{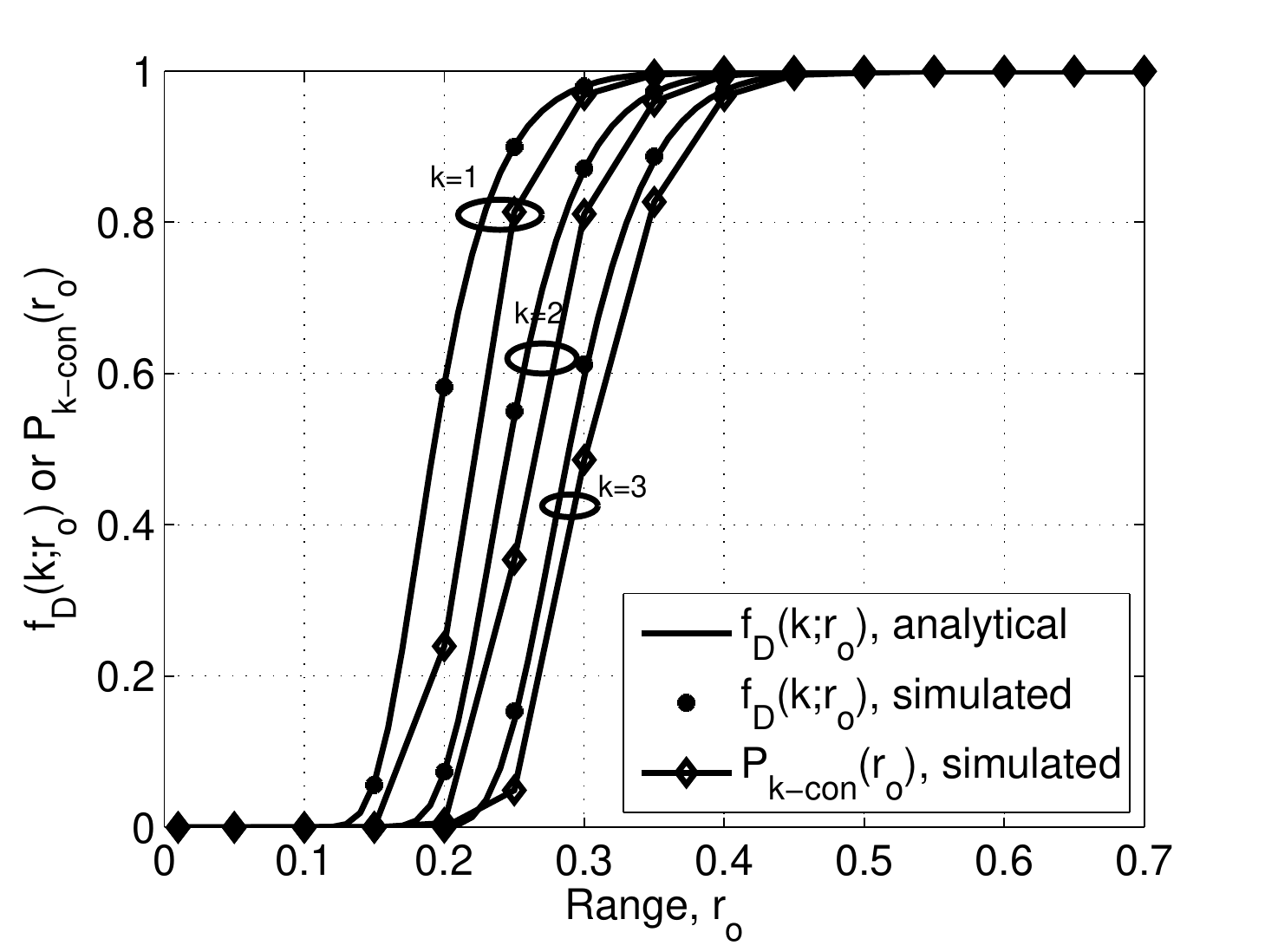}}
     \caption{Minimum node degree distribution, $f_D(k;\rr)$ and probability of $k$-connectivity, $\pcon$, versus transmission range $\rr$ for (a) $N=10$, (b) $N=20$ and (c) $N=50$ nodes independently and uniformly distributed in a unit square.}
    \label{fig:node_degree}
\end{figure*}

Here, through simulations, we validate our framework to determine
the minimum node degree distribution. Using
\eqref{Eq:min_node_degree_sum}, we determine the minimum node degree
distribution $f_D(k;\rr)$ for $k=1, 2, 3$ and for number of nodes
$N=10, 20, 50$. The simulation results for $f_D(k;\rr)$ are obtained
by uniformly distributing the $N$ nodes in a square region, each
with transmission range $\rr$, over a square region and determining
if the minimum of number of neighbors for all nodes in the network
are equal to $k$. Note that the simulation results for $f_D(k;\rr)$
are obtained by averaging over all $S=50,000$ random topologies for
each $\rr$ and $k$ . The simulation results for $f_D(k;\rr)$ are
also plotted in \figref{fig:node_degree} and match perfectly with
the analytical results using \eqref{Eq:min_node_degree_sum}.


As highlighted earlier in Section II-B, the probability of
$k$-connectivity $\pkcon$ is bounded by the minimum node degree
distribution $f_D(k;\rr)$. This is because we obtain a $k$-connected
network at the same time when we obtain a network with minimum node
degree $k$, both with and without boundary
effects~\cite{Penrose-1997,bettstetter-2002,Bettstetter:2004}. Here
we validate through simulation results that the minimum node degree
distribution $f_D(k;\rr)$ serves as an upper bound for $\pkcon$ and
the bound gets tighter as $\pkcon$ approaches one.

We repeat the simulation environment of
Section~\ref{sec:results_min_node} and now for each of the $50,000$
random topologies, we measure the $k$-connectivity of the network
for $k = 1, 2, 3$. The simulation results for $\pkcon$ are plotted
in \figref{fig:node_degree} along with the analytical results for
minimum node degree distribution $f_D(k;\rr)$ obtained using
\eqref{Eq:min_node_degree_sum} and our proposed framework. It is
evident in the plots that the relation between $f_D(k;\rr)$ and
$\pkcon$ given in \eqref{Eq:pkcon_bound} holds, that is, $f_D(k;\rr)
\rightarrow \pkcon$ as $\pkcon \rightarrow 1$.

We note that the simulation tests for $k$-connectivity are
computationally intensive and the computational complexity to check
$k$-connectivity scales with $N^{k-1}$ for $k\geq2$. For example,
the complexity to check $1$-connectivity and $2$-connectivity is of
the order $O(N+S)$ and the complexity to determine $3$-connectivity
is $O(N(N + S))$, where $S$ denotes the number of simulations.
However, \eqref{Eq:min_node_degree_sum} can be evaluated numerically
very easily. This illustrates an advantage of the proposed framework
over simulations.

%
\subsection{$1$-Connectivity}

Since $\pcon$ is one of the essential characteristics of wireless
multi-hop networks, we analyze the tightness of the bound provided
by $f_D(1;\rr)$ for $\pcon$ in more detail and compare with the
existing bounds in the literature.

\begin{figure}[t]
        \vspace{-3mm}
    \centering
    \includegraphics[width=0.70\textwidth]{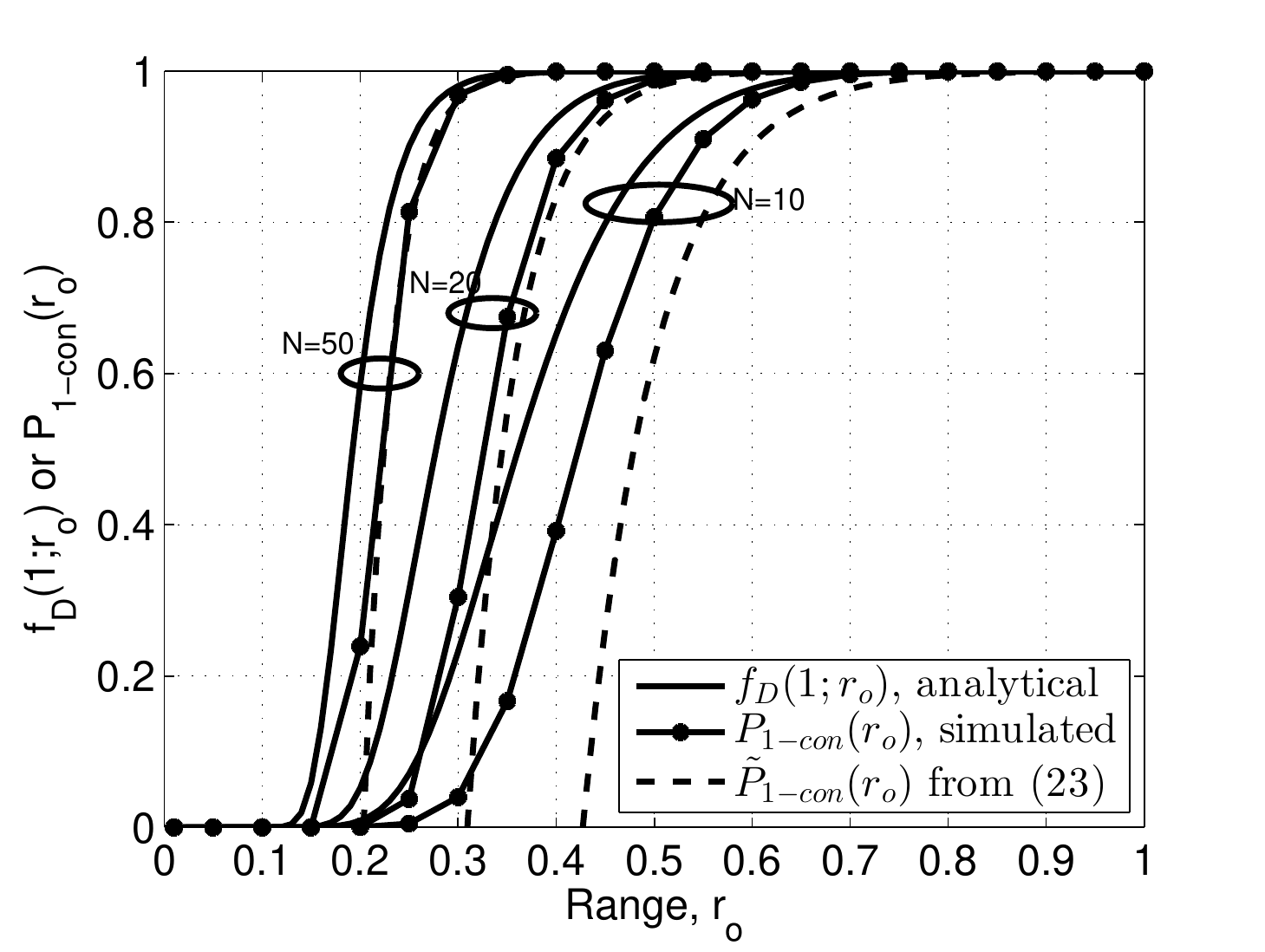}
     \caption{Probability of connectivity: Upper bound as minimum node degree distribution, $f_D(k;\rr)$, high density approximation $\tilde P_{\textrm{1-con}}(\rr)$ and simulated $\pcon$, versus transmission range $\rr$ for $N=10,20,50$ nodes independently and uniformly distributed in a unit square.}
    \label{fig:nodes_con}
\end{figure}

\figref{fig:nodes_con} plots the $f_D(1;\rr)$ as an upper bound for
the probability of connectivity (obtained via simulations), versus
transmission range $\rr$ for $N=10,20, 50$ nodes. For comparison, we
plot the high density approximation for $\pcon$ which is derived
in~\cite{Coon-2012} using a cluster expansion approach as
\begin{align}\label{eq:coon}
\tilde P_{\textrm{1-con}}(\rr) &\approx 1- L^2\rho
e^{-\frac{\pi}{\beta}\rho}-4L
\sqrt{\frac{\beta}{\pi}}e^{-\frac{\pi}{2\beta}\rho}- \frac{16
\beta}{\rho \pi} e^{-\frac{\pi}{4\beta}\rho},
\end{align}

\noindent where $L$ denotes the side length, $\rho$ denotes the node
density and $\beta = (\rr/L)^{-2}$.


\begin{figure*}[t]
    \vspace{-3mm}
    \centering
    \subfloat[$N=10$]
        {\label{fig:nodes_con_diff_a}\includegraphics[scale=0.42]{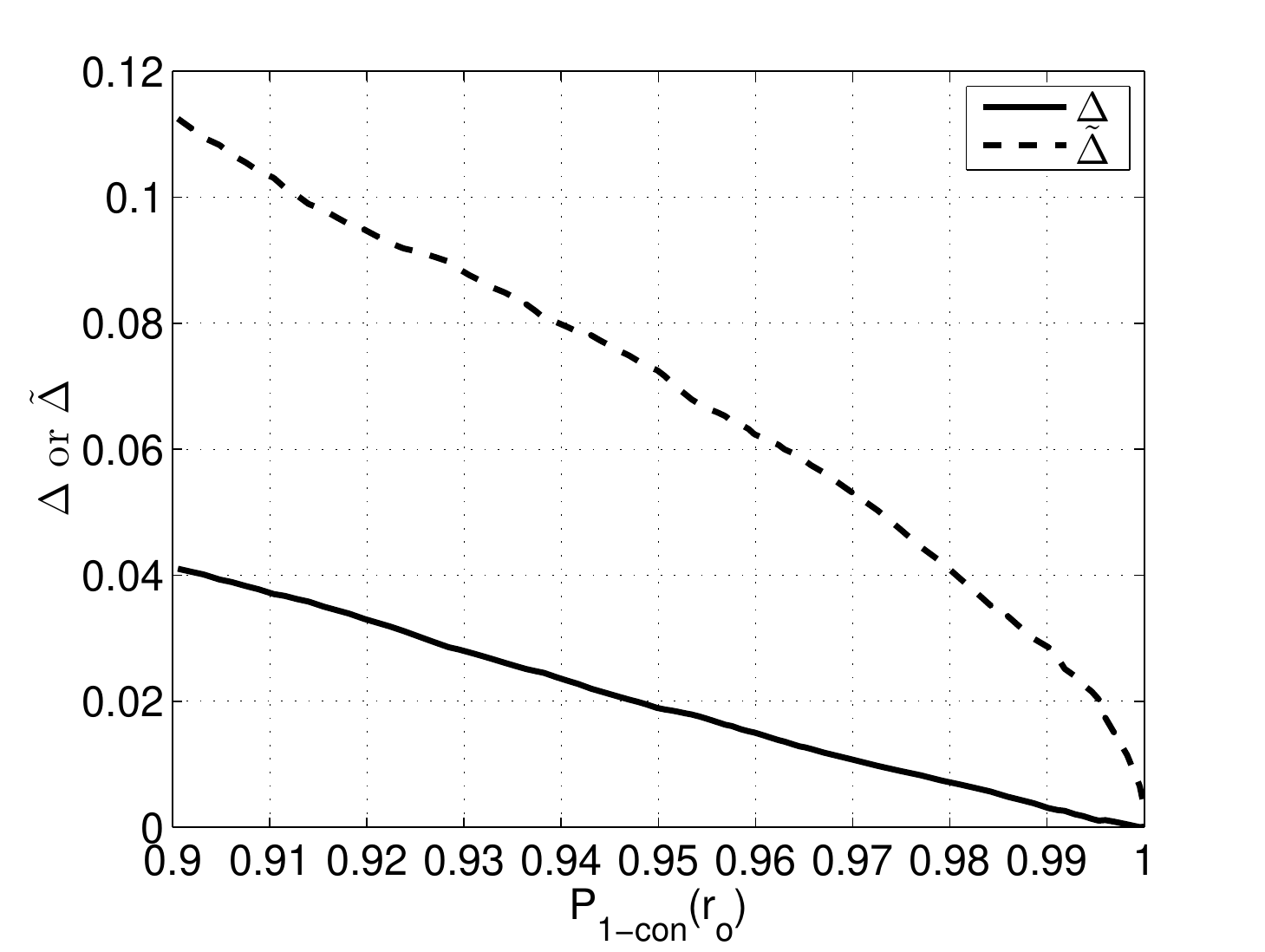}}
    \subfloat[$N=20$]
        {\label{fig:nodes_con_diff_b}\includegraphics[scale=0.42]{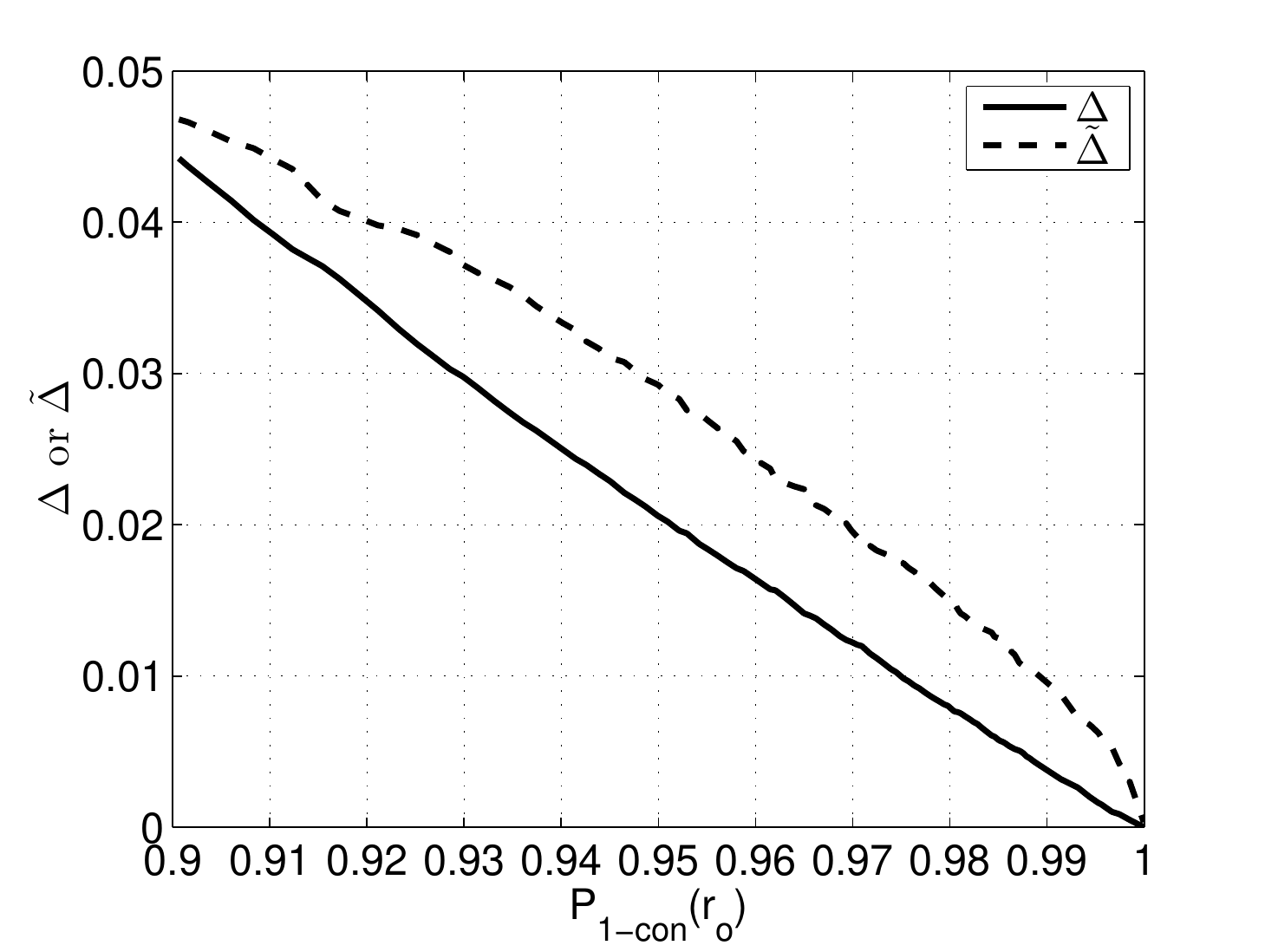}}
    \subfloat[$N=50$]
        {\label{fig:nodes_con_diff_c}\includegraphics[scale=0.42]{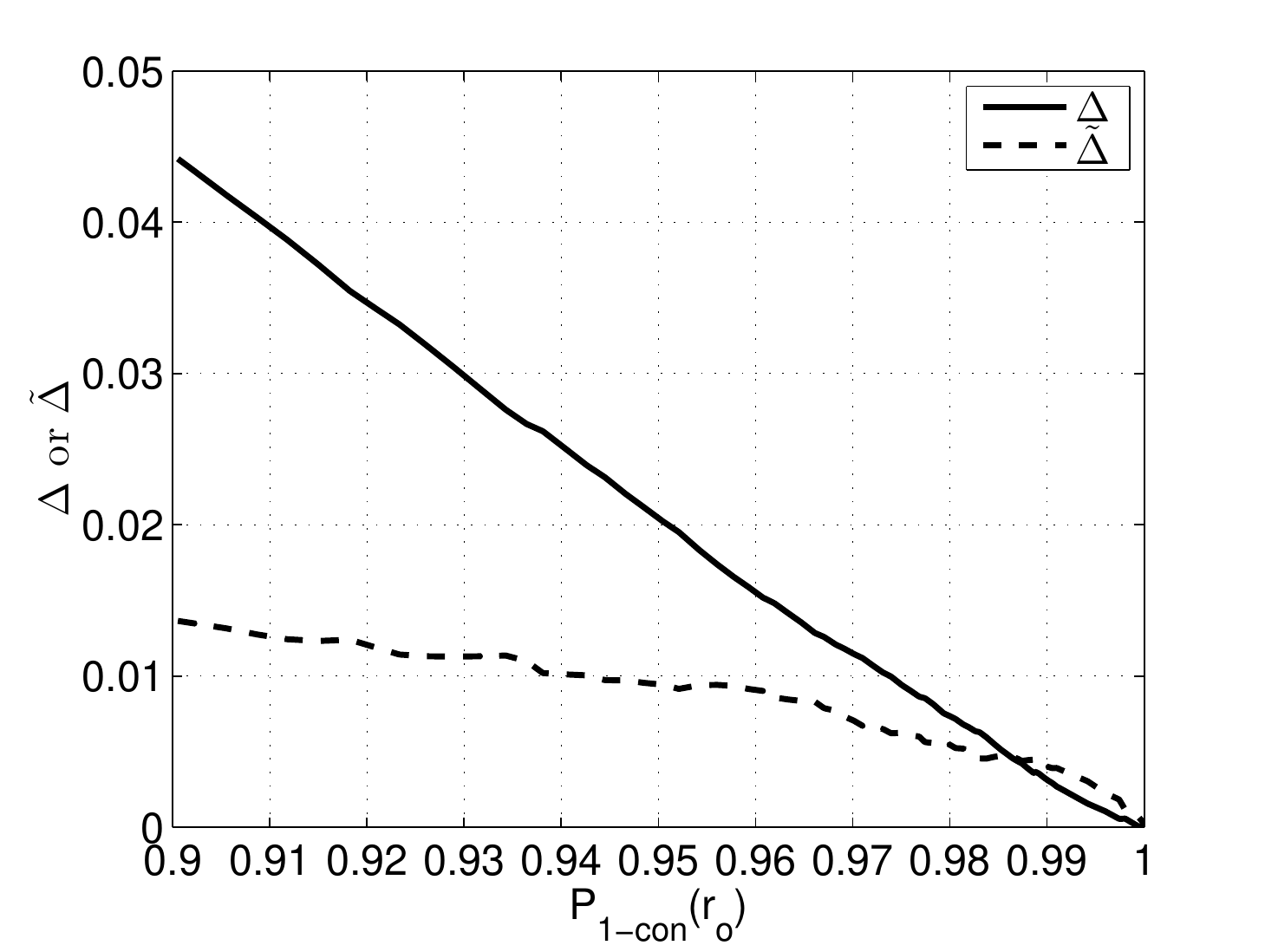}}
     \caption{The differences $\Delta=
|{P_{\textrm{1-con}}(\rr)} -f_D(1;\rr)|$ and $\tilde{\Delta} =
|{P_{\textrm{1-con}}(\rr)} -\tilde P_{\textrm{1-con}}(\rr)|$ denotes
the deviation of simulated $\pcon$ from the proposed bound as
minimum node degree distribution $f_D(1;\rr)$ in
\eqref{Eq:min_node_degree_sum} and high density approximation
$\tilde P_{\textrm{1-con}}(\rr)$ given in \eqref{eq:coon}
respectively. Both $\Delta$ and $\tilde \Delta$ are plotted for (a)
$N=10$, (b) $N=20$ and (c) $N=50$.}
    \label{fig:nodes_con_diff}
\end{figure*}

The high density approximation in \eqref{eq:coon} is comparatively a
better estimate for $\pcon$ for $N=50$ nodes but is not useful for
$N=10$ nodes. We analyze the tightness of the bounds in more detail over
the value of probabilities in the interval $0.9 \leq \pcon \leq 1$.
The differences between the bounds, $\Delta=
|{P_{\textrm{1-con}}(\rr)} -f_D(1;\rr)|$ and $\tilde{\Delta} =
|{P_{\textrm{1-con}}(\rr)} -\tilde P_{\textrm{1-con}}(\rr)|$ is
plotted in \figref{fig:nodes_con_diff} for $0.9\leq
P_{\textrm{1-con}}(\rr) \leq 1$. For small number of
nodes~($N=10,20$), it can be noted that the proposed minimum node
degree distribution $f_D(1;\rr)$ is comparatively a tight bound for
$\pcon$ than the high density approximation $\tilde
P_{\textrm{1-con}}(\rr)$.

On the basis of simulation results presented here and in the
previous section, we can say that the upper bound for the connectivity
provided by the minimum node degree distribution in
\eqref{Eq:pkcon_bound} provides a good approximation for the
simulation results when $\pcon \approx 1$. This is consistent with
the observation in~\cite{Bettstetter:2004} for circular areas with
or without boundary effects. Thus, the proposed framework can be
used to accurately predict the network connectivity properties even
when the number of nodes is small.

\subsection{Network Design Example: Minimum Transmission Range and Minimum Number of Nodes}

We now address the network design problems: a) Determine the minimum
transmission range $r_0^c$ for given number of nodes or b) Find are
the minimum number of nodes $N^c$, each with given transmission
range $\rr$, such that the network is $k$-connected with high
probability $\pkcon$, say 0.95 or 0.99. Such a minimum value
transmission range and the minimum number of nodes to achieve the
desired level of $\pkcon$ are often termed as critical transmission
range and critical number of nodes
respectively~\cite{bettstetter-2002,Peng:2005,Peng:2010}. For a given
transmission range, the minimum number of nodes must be deployed to
minimize the cost and reduce the interference between the
nodes~\cite{bettstetter-2002}. For a network with fixed number of
nodes, the transmission range must be large enough to ensure the
network connectivity but it should be small enough to minimize the
power consumption and reduce the interference between the nodes.

\begin{figure}[t]
    \vspace{-3mm}
    \centering
    \includegraphics[width=0.70\textwidth]{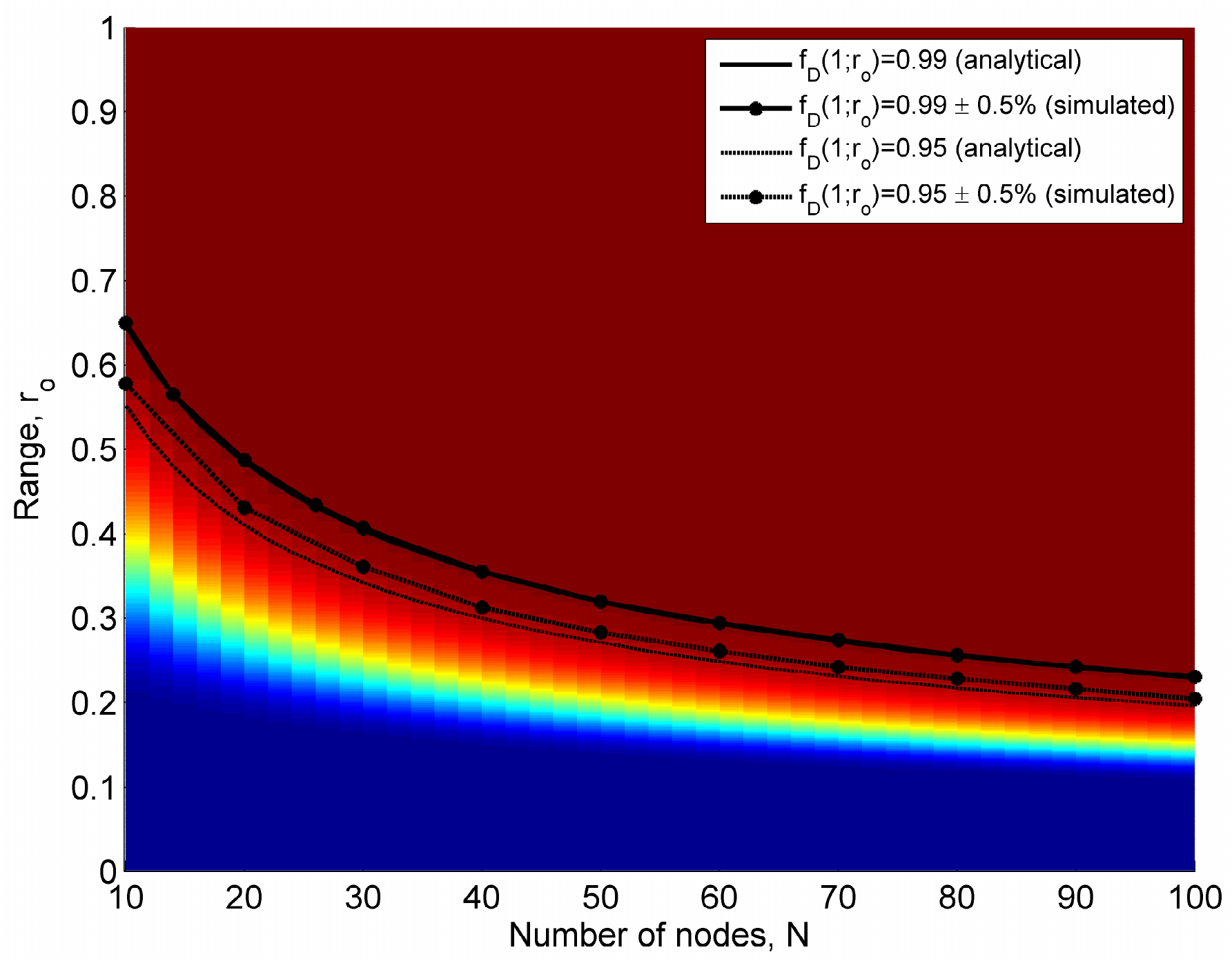}
     \caption{Surface plot for $f_D(1;\rr)$ as a
function of number of nodes $N$ and transmission range $\rr$. The
curves denote $(N,\rr)$ pairs for which $f_D(1;\rr)=0.95,$ or $0.99$
and $(N^c,\rr^c)$ pairs for which $\pkcon=0.95\pm 0.5\%$ or $0.99\pm
0.5\%$. }
    \label{fig:critical_range_nodes}
\end{figure}

Since $\pkcon$ is a monotonic function of both transmission range
$\rr$ and the number of nodes $N$, the solution to the above problem is to
determine the curve in $N$-$\rr$ plane for which $\pkcon=0.95$ or
$\pkcon=0.99$. We can carry out simulations to determine $\rr^c$ or
$N^c$. However, as highlighted earlier, it would be very
computationally intensive to obtain simulation results with
sufficient accuracy, especially for large values of $k$. We
demonstrate here that we can obtain analytical solutions to the
design problems mentioned above using the proposed framework.
Recalling that minimum node degree distribution $f_D(k;\rr)$ serves
as a good approximation for $\pkcon$, we can use $f_D(k;\rr)$ given
in \eqref{Eq:min_node_degree_sum} to determine the critical
transmission range and critical number of nodes. The surface plot for
$f_D(1;\rr)$ is shown in \figref{fig:critical_range_nodes} as a
function of the number of nodes $N$ and the transmission range $\rr$, where
we have also shown the analytical curves consisting of $(N,\rr)$
pairs for which $f_D(1;\rr) = 0.95$ or $f_D(1;\rr) = 0.99$ and the
simulation curves denoting $(N^c,\rr^c)$ pairs for which $\pcon =
0.95$ or $\pcon = 0.99$ with the tolerance of $\pm 0.5\%$. It can be
observed that the analytical determination of $(N^c,\rr^c)$ using
the proposed minimum node degree distribution $f_D(1;\rr)$ yields a
fairly good approximation for both $\pcon=0.95$ and $\pcon=0.99$.
Nevertheless, by virtue of our analysis in the previous section and
Penrose theorem on connectivity of random
graphs~\cite{Penrose-1997}, the analytical determination is more
accurate for $\pcon=0.99$.


\section{Conclusions}\label{sec:conclusions}
In this paper, we have presented a tractable analytical framework
for the exact calculation of the probability of node isolation and
the minimum node degree distribution in finite wireless sensor networks.
We have considered $N$ sensor nodes, each with transmission range
$\rr$, which are independently and uniformly distributed in a square
region. The proposed framework can accurately account for the
boundary effects by partitioning the square into subregions, based
on the transmission range and the node location. The exact modeling
of the boundary effects has not been taken into consideration in
previous studies in the literature. Our results confirm that the
boundary effects play a key role in determining the connectivity
metrics of the network: probability of node isolation, minimum node
degree distribution and probability of $k$-connectivity, especially
when the number of nodes is small $(N < 50$). We have also validated
the proposed framework with the help of simulations.

Future research can consider natural generalizations of the work
presented here. Firstly, the proposed framework can be extended for
the case of polygon region (generalization of square region).
Secondly, the consideration of the channel fading and interference
in the transmission model~(generalization of disk model) is an open
problem which is outside the scope of this work.

\begin{appendices}
\section{Minimum Node Degree Distribution}
Here, we present the formulation of the probability distribution of
the minimum node degree $D$ presented in \eqref{Eq:min_node_degree}.
For $N$ uniformly distributed nodes, the number of neighbors $d$ for
a node located at $\dv{u}$ follows a binomial
distribution~\cite{bettstetter-2002,Bettstetter:2004,Srinivasa-2010}
\[
\binom{N-1}{d}  \left( F(\dv{u};\rr) \right)^{d} \left(1-
F(\dv{u};\rr) \right)^{N-d-1}
\]
and the probability that \emph{any} node in the network has at least
$d$ neighbors is therefore given by
\[
\binom{N-1}{d} \int_\mathcal{R} \left( F(\dv{u};\rr) \right)^{d}
\left(1- F(\dv{u};\rr) \right)^{N-d-1} ds(\dv{u}).
\]
Now the probability that any node in the network has at least $k$
neighbors can be expressed as
\[
1- \sum_{d=0}^{k-1}\binom{N-1}{d} \int_\mathcal{R} \left(
F(\dv{u};\rr) \right)^{d} \left(1- F(\dv{u};\rr) \right)^{N-d-1}
ds(\dv{u}),
\]
which gives rise to the minimum node degree distribution $f_D(k;
\rr) = \textrm{P}(D=k)$ that all the nodes have at least $k$
neighbors, with an assumption of independence between the nodes.
This leads to the result in \eqref{Eq:min_node_degree}.
\end{appendices}

\ifCLASSOPTIONonecolumn
    \renewcommand{\baselinestretch}{1.97}
\fi



\end{document}